\documentclass[12pt,reqno]{amsart}
\usepackage{graphicx}
\usepackage{amscd}
\usepackage{amsmath}
\usepackage{epsfig}
\usepackage{amsfonts}
\usepackage{amssymb}

\providecommand{\U}[1]{\protect\rule{.1in}{.1in}}
\providecommand{\U}[1]{\protect\rule{.1in}{.1in}}
\allowdisplaybreaks

\theoremstyle{plain}

\numberwithin{equation}{section}

\input{tcilatex}

\begin{document}
\title[Schr\"{o}dinger Equation and Airy Functions]{The Time-Dependent Schr%
\"{o}dinger Equation, \\
Riccati Equation and Airy Functions}
\author{Nathan Lanfear}
\address{School of Mathematical and Statistical Sciences, Arizona State
University, Tempe, AZ 85287--1804, U.S.A.}
\email{nlanfear@asu.edu}
\author{Sergei K. Suslov}
\address{School of Mathematical and Statistical Sciences, Arizona State
University, Tempe, AZ 85287--1804, U.S.A.}
\email{sks@asu.edu}
\urladdr{http://hahn.la.asu.edu/\symbol{126}suslov/index.html}
\dedicatory{Dedicated to Dick Askey on his 75th birthday}
\date{\today }
\subjclass{Primary 81Q05, 35C05. Secondary 42A38}
\keywords{The time-dependent Schr\"{o}dinger equation, Cauchy initial value
problem, Riccati differential equation, Green function, propagator, gauge
transformation, nonlinear Schr\"{o}dinger equation, quantum parametric
oscillator, Meixner polynomials, Bargmann's functions}

\begin{abstract}
We construct the Green functions (or Feynman's propagators) for the Schr\"{o}%
dinger equations of the form $i\psi _{t}+\frac{1}{4}\psi _{xx}\pm tx^{2}\psi
=0$ in terms of Airy functions and solve the Cauchy initial value problem in
the coordinate and momentum representations. Particular solutions of the
corresponding nonlinear Schr\"{o}dinger equations with variable coefficients
are also found. A special case of the quantum parametric oscillator is
studied in detail first. The Green function is explicitly given in terms of
Airy functions and the corresponding transition amplitudes are found in
terms of a hypergeometric function. The general case of quantum parametric
oscillator is considered then in a similar fashion. A group theoretical
meaning of the transition amplitudes and their relation with Bargmann's
functions is established.
\end{abstract}

\maketitle

\section{Introduction}

In this paper we discuss explicit solutions of the Cauchy initial value
problem for the one-dimensional Schr\"{o}dinger equations%
\begin{equation}
i\frac{\partial \psi }{\partial t}+\frac{1}{4}\frac{\partial ^{2}\psi }{%
\partial x^{2}}\pm tx^{2}\psi =0,\qquad \psi \left( x,0\right) =\varphi
\left( x\right)  \label{in1}
\end{equation}%
with a suitable initial data on the entire real line $\boldsymbol{R}.$ The
corresponding Green functions are found in terms of compositions of
elementary and Airy functions in the coordinate and momentum
representations. It is well-known that the Airy equation describes motion of
a quantum particle in the neighborhood of the turning point\ on the basis of
the stationary, or time-independent, Schr\"{o}dinger equation \cite%
{Brillouin}, \cite{Kramers}, \cite{Wentzel}, \cite{La:Lif}, and \cite{Merz}.
Here we consider an application of these functions to the time-dependent Schr%
\"{o}dinger equations for certain parametric oscillator.

It is worth noting that the Green functions for the Schr\"{o}dinger equation
are known explicitly only in a few special cases. An important example of
this source is the forced harmonic oscillator originally considered by
Richard Feynman in his path integrals approach to the nonrelativistic
quantum mechanics \cite{FeynmanPhD}, \cite{Feynman}, \cite{Feynman49a}, \cite%
{Feynman49b}, and \cite{Fey:Hib}; see also \cite{Lop:Sus}. Since then this
problem and its special and limiting cases were discussed by many authors;
see Refs.~\cite{Beauregard}, \cite{Gottf:T-MY}, \cite{Holstein}, \cite%
{Maslov:Fedoriuk}, \cite{Merz}, \cite{Thomber:Taylor} for the simple
harmonic oscillator and Refs.~\cite{Arrighini:Durante}, \cite{Brown:Zhang}, 
\cite{Holstein97}, \cite{Nardone}, \cite{Robinett} for the particle in a
constant external field and references therein.

The case of Schr\"{o}dinger equation with a general variable quadratic
Hamiltonian is investigated in Ref.~\cite{Cor-Sot:Lop:Sua:Sus}; see also 
\cite{Cor-Sot:Sua:Sus}, \cite{Cor-Sot:Sus}, \cite{Lop:Sus}, \cite{Me:Co:Su}, 
\cite{Sua:Sus}, and \cite{Suaz:Sus}. Here we present a few examples that are
integrable in terms of Airy functions. In this approach, all known exactly
solvable quadratic models are classified in terms of solutions of a certain
characterization equation. These exactly solvable cases may be of interest
in a general treatment of the linear and nonlinear evolution equations; see 
\cite{BatemanPDE}, \cite{Cann}, \cite{Caz}, \cite{Caz:Har}, \cite%
{Lad:Sol:Ural}, \cite{Levi}, \cite{Meln:Filin}, \cite{Sua:Sus:Vega}, \cite%
{Tao} and references therein. Moreover, these explicit solutions can also be
useful when testing numerical methods of solving the time-dependent Schr\"{o}%
dinger equations with variable coefficients. The solution of the quantum
parametric oscillator problem found in this paper is also relevant.

\section{Green Function: Increasing Case}

The fundamental solution of the time-dependent Schr\"{o}dinger equation%
\begin{equation}
i\frac{\partial \psi }{\partial t}+\frac{1}{4}\frac{\partial ^{2}\psi }{%
\partial x^{2}}+tx^{2}\psi =0  \label{dec1}
\end{equation}%
can be found by a familiar substitution \cite{Cor-Sot:Lop:Sua:Sus}%
\begin{equation}
\psi =A\left( t\right) e^{iS\left( x,y,t\right) }=\frac{1}{\sqrt{2\pi i\mu
\left( t\right) }}e^{i\left( \alpha \left( t\right) x^{2}+\beta \left(
t\right) xy+\gamma \left( t\right) y^{2}\right) }.  \label{dec2}
\end{equation}%
The real-valued functions of time $\alpha \left( t\right) ,$ $\beta \left(
t\right) ,$ $\gamma \left( t\right) $ satisfy the following system of
ordinary differential equations%
\begin{eqnarray}
&&\frac{d\alpha }{dt}-t+\alpha ^{2}=0,  \label{dec3} \\
&&\frac{d\beta }{dt}+\alpha \beta =0,  \label{dec4} \\
&&\frac{d\gamma }{dt}+\frac{1}{4}\beta ^{2}=0,  \label{dec5}
\end{eqnarray}%
where the first equation is the special Riccati nonlinear differential
equation; see, for example, \cite{Fra:Gar:Frag}, \cite{Haah:Stein}, \cite%
{Molch}, \cite{Rainville}, \cite{Rajah:Mah}, \cite{Wa} and references
therein.

The substitution 
\begin{equation}
\alpha =\frac{\mu ^{\prime }}{\mu },\qquad \alpha ^{\prime }=\frac{\mu
^{\prime \prime }}{\mu }-\left( \frac{\mu ^{\prime }}{\mu }\right) ^{2},
\label{dec6}
\end{equation}%
which according to Ref.~\cite{Molch} goes back to Jean le Round d'Alembert 
\cite{D'Alembert}, results in the second order linear equation%
\begin{equation}
\mu ^{\prime \prime }-t\mu =0.  \label{dec7}
\end{equation}%
The initial conditions for the corresponding Green function are $\mu \left(
0\right) =0$ and $\mu ^{\prime }\left( 0\right) =1/2.$ It is well-known that
Eq.~(\ref{dec7}) can be solved in terms of Airy functions which are studied
in detail; see, for example, \cite{Ab:St}, \cite{An:As:Ro}, \cite{Molch}, 
\cite{Ni:Uv}, \cite{Sus:Trey}, \cite{Wa} and references therein. A different
definition of these functions that is convenient for our purposes in this
paper is given in the Appendix~A.

We choose $\mu _{0}=\left( 1/2\right) a\left( t\right) $ and the required
Green function solution of the system is given by%
\begin{equation}
\alpha _{0}=\frac{a^{\prime }\left( t\right) }{a\left( t\right) },\qquad
\beta _{0}=-\frac{2}{a\left( t\right) },\qquad \gamma _{0}=\frac{b\left(
t\right) }{a\left( t\right) },  \label{dec8}
\end{equation}%
where the Airy functions $a\left( t\right) =ai\left( t\right) $ and $b\left(
t\right) =bi\left( t\right) $ are defined by (\ref{airy12}) and (\ref{airy13}%
), respectively. Indeed,%
\begin{equation}
\frac{d\beta _{0}}{dt}=-2\left( a^{-1}\right) ^{\prime }=2\frac{a^{\prime }}{%
a^{2}}=-\alpha _{0}\beta _{0},  \label{dec9}
\end{equation}%
and%
\begin{equation}
\frac{d\gamma _{0}}{dt}=\left( \frac{b}{a}\right) ^{\prime }=\frac{b^{\prime
}a-ba^{\prime }}{a^{2}}=\frac{W\left( a,b\right) }{a^{2}}=-\frac{1}{4}\beta
_{0}^{2}.  \label{dec10}
\end{equation}%
Thus, the Green function has the following closed form%
\begin{equation}
G\left( x,y,t\right) =\frac{1}{\sqrt{\pi ia\left( t\right) }}\exp \left( i%
\frac{a^{\prime }\left( t\right) x^{2}-2xy+b\left( t\right) y^{2}}{a\left(
t\right) }\right) ,\qquad t>0  \label{dec11}
\end{equation}%
in terms of elementary and Airy functions.

It is worth noting that a more general particular solution has the form%
\begin{equation}
\psi =K\left( x,y,t\right) =\frac{1}{\sqrt{2\pi i\mu \left( t\right) }}\
e^{i\left( \alpha \left( t\right) x^{2}+\beta \left( t\right) xy+\gamma
\left( t\right) y^{2}\right) },  \label{dec13}
\end{equation}%
where $\mu =c_{1}a\left( t\right) +c_{2}b\left( t\right) $ with $\mu \left(
0\right) =c_{2}\neq 0,$ $\mu ^{\prime }\left( 0\right) =c_{1}$ and%
\begin{eqnarray}
&&\alpha =\frac{c_{1}a^{\prime }\left( t\right) +c_{2}b^{\prime }\left(
t\right) }{c_{1}a\left( t\right) +c_{2}b\left( t\right) },\qquad \alpha
\left( 0\right) =\frac{c_{1}}{c_{2}},  \label{dec14} \\
&&\beta =\frac{c_{2}\beta \left( 0\right) }{c_{1}a\left( t\right)
+c_{2}b\left( t\right) },  \label{dec15} \\
&&\gamma =\gamma \left( 0\right) -\frac{c_{2}\beta ^{2}\left( 0\right)
a\left( t\right) }{4\left( c_{1}a\left( t\right) +c_{2}b\left( t\right)
\right) }.  \label{dec16}
\end{eqnarray}%
This can be easily verified by a direct substitution into the system (\ref%
{dec3})--(\ref{dec5}).

\section{Initial Value Problem: Increasing Case}

The solution of the Cauchy initial value problem%
\begin{equation}
i\frac{\partial \psi }{\partial t}+\frac{1}{4}\frac{\partial ^{2}\psi }{%
\partial x^{2}}+tx^{2}\psi =0,\qquad \psi \left( x,0\right) =\varphi \left(
x\right)  \label{dec11a}
\end{equation}%
is given by the superposition principle in an integral form%
\begin{equation}
\psi \left( x,t\right) =\int_{-\infty }^{\infty }G\left( x,y,t\right) \
\varphi \left( y\right) \ dy,  \label{dec12}
\end{equation}%
where one should justify interchange differentiation and integration for a
suitable initial function $\varphi $ on $\boldsymbol{R};$ a rigorous proof
is given in Ref.~\cite{Suaz:Sus}.

The special case $\varphi \left( y\right) =K\left( z,y,0\right) $ of the
time evolution operator (\ref{dec12}) is%
\begin{equation}
K\left( x,y,t\right) =\int_{-\infty }^{\infty }G\left( x,z,t\right) \
K\left( z,y,0\right) \ dz  \label{dec25}
\end{equation}%
and its inversion is given by%
\begin{equation}
G\left( x,y,t\right) =\mu \left( 0\right) \left\vert \beta \left( 0\right)
\right\vert \int_{-\infty }^{\infty }K\left( x,z,t\right) \ K^{\ast }\left(
y,z,0\right) \ dz,  \label{dec26}
\end{equation}%
where the star denotes the complex conjugate. The familiar
Euler--Gaussian--Fresnel integral \cite{Bo:Shi} and \cite{Palio:Mead},%
\begin{equation}
\int_{-\infty }^{\infty }e^{i\left( az^{2}+2bz\right) }\,dz=\sqrt{\frac{\pi i%
}{a}}\,e^{-ib^{2}/a},\qquad \func{Im}a\geq 0,  \label{gauss}
\end{equation}%
allows to obtain the following transformation \cite{Suaz:Sus}%
\begin{eqnarray}
&&\mu \left( t\right) =2\mu \left( 0\right) \mu _{0}\left( t\right) \left(
\alpha \left( 0\right) +\gamma _{0}\left( t\right) \right) ,  \label{dec17}
\\
&&\alpha \left( t\right) =\alpha _{0}\left( t\right) -\frac{\beta
_{0}^{2}\left( t\right) }{4\left( \alpha \left( 0\right) +\gamma _{0}\left(
t\right) \right) },  \label{dec18} \\
&&\beta \left( t\right) =-\frac{\beta \left( 0\right) \beta _{0}\left(
t\right) }{2\left( \alpha \left( 0\right) +\gamma _{0}\left( t\right)
\right) },  \label{dec19} \\
&&\gamma \left( t\right) =\gamma \left( 0\right) -\frac{\beta ^{2}\left(
0\right) }{4\left( \alpha \left( 0\right) +\gamma _{0}\left( t\right)
\right) }  \label{dec20}
\end{eqnarray}%
and its inverse%
\begin{eqnarray}
&&\mu _{0}\left( t\right) =\frac{2\mu \left( t\right) }{\mu \left( 0\right)
\beta ^{2}\left( 0\right) }\left( \gamma \left( 0\right) -\gamma \left(
t\right) \right) ,  \label{dec21} \\
&&\alpha _{0}\left( t\right) =\alpha \left( t\right) +\frac{\beta ^{2}\left(
t\right) }{4\left( \gamma \left( 0\right) -\gamma \left( t\right) \right) },
\label{dec22} \\
&&\beta _{0}\left( t\right) =-\frac{\beta \left( 0\right) \beta \left(
t\right) }{2\left( \gamma \left( 0\right) -\gamma \left( t\right) \right) },
\label{dec23} \\
&&\gamma _{0}\left( t\right) =-\alpha \left( 0\right) +\frac{\beta
^{2}\left( 0\right) }{4\left( \gamma \left( 0\right) -\gamma \left( t\right)
\right) }  \label{dec24}
\end{eqnarray}%
in the cases (\ref{dec25}) and (\ref{dec26}), respectively. Direct
calculation shows, once again, that our solutions (\ref{dec8}) and (\ref%
{dec14})--(\ref{dec16}) do satisfy these transformation rules. It is worth
noting that the transformation (\ref{dec21})--(\ref{dec24}) allows to derive
our Green function from any regular solution of the system (\ref{dec3})--(%
\ref{dec5}).

\section{Oscillatory Case}

A time-dependent Schr\"{o}dinger equation%
\begin{equation}
i\frac{\partial \psi }{\partial t}+\frac{1}{4}\frac{\partial ^{2}\psi }{%
\partial x^{2}}-tx^{2}\psi =0  \label{osc1}
\end{equation}%
can be solved in a similar fashion by the substitution (\ref{dec2}) with%
\begin{eqnarray}
&&\frac{d\alpha }{dt}+t+\alpha ^{2}=0,  \label{osc2} \\
&&\frac{d\beta }{dt}+\alpha \beta =0,  \label{osc3} \\
&&\frac{d\gamma }{dt}+\frac{1}{4}\beta ^{2}=0.  \label{osc4}
\end{eqnarray}%
Here $\mu _{0}=-\left( 1/2\right) a\left( -t\right) $ and%
\begin{equation}
\alpha _{0}=-\frac{a^{\prime }\left( -t\right) }{a\left( -t\right) },\qquad
\beta _{0}=\frac{2}{a\left( -t\right) },\qquad \gamma _{0}=-\frac{b\left(
-t\right) }{a\left( -t\right) }.  \label{osc5}
\end{equation}%
The Green function is%
\begin{equation}
G\left( x,y,t\right) =\frac{1}{\sqrt{-\pi ia\left( -t\right) }}\exp \left( -i%
\frac{a^{\prime }\left( -t\right) -2xy+b\left( -t\right) y^{2}}{a\left(
-t\right) }\right) ,\qquad t>0  \label{osc6}
\end{equation}%
and the solution of the initial value problem is given by the integral (\ref%
{dec12}).

A more general particular solution has the form (\ref{dec13}), where $\mu
=c_{1}a\left( -t\right) +c_{2}b\left( -t\right) $ with $\mu \left( 0\right)
=c_{2}\neq 0,$ $\mu ^{\prime }\left( 0\right) =-c_{1}$ and%
\begin{eqnarray}
&&\alpha =-\frac{c_{1}a^{\prime }\left( -t\right) +c_{2}b^{\prime }\left(
-t\right) }{c_{1}a\left( -t\right) +c_{2}b\left( -t\right) },\qquad \alpha
\left( 0\right) =-\frac{c_{1}}{c_{2}},  \label{osc7} \\
&&\beta =\frac{c_{2}\beta \left( 0\right) }{c_{1}a\left( -t\right)
+c_{2}b\left( -t\right) },  \label{osc8} \\
&&\gamma =\gamma \left( 0\right) -\frac{c_{2}\beta ^{2}\left( 0\right)
a\left( -t\right) }{4\left( c_{1}a\left( -t\right) +c_{2}b\left( -t\right)
\right) }.  \label{osc9}
\end{eqnarray}%
This can be easily verified by a direct substitution into the system (\ref%
{osc2})--(\ref{osc4}) or with the aid of the transformations (\ref{dec17})--(%
\ref{dec20}) and (\ref{dec21})--(\ref{dec24}). We leave further details to
the reader.

\section{Momentum Representation}

The Schr\"{o}dinger equation (\ref{dec1}) takes the form%
\begin{equation}
i\frac{\partial \psi }{\partial t}-t\frac{\partial ^{2}\psi }{\partial x^{2}}%
-\frac{1}{4}x^{2}\psi =0  \label{mom1}
\end{equation}%
in the momentum representation; see, for example, Ref.~\cite{Cor-Sot:Sus}
for more details. The substitution (\ref{dec2}) results in%
\begin{eqnarray}
&&\frac{d\alpha }{dt}+\frac{1}{4}-4t\alpha ^{2}=0,  \label{mom2} \\
&&\frac{d\beta }{dt}-4t\alpha \beta =0,  \label{mom3} \\
&&\frac{d\gamma }{dt}-t\beta ^{2}=0.  \label{mom4}
\end{eqnarray}%
The Riccati equation (\ref{mom2}) by the standard substitution%
\begin{equation}
\alpha =-\frac{1}{4t}\frac{\mu ^{\prime }}{\mu }  \label{mom5}
\end{equation}%
is transformed to the second order linear equation%
\begin{equation}
\mu ^{\prime \prime }-\frac{1}{t}\mu ^{\prime }-t\mu =0,  \label{mom6}
\end{equation}%
whose linearly independent solutions are the derivatives of Airy functions $%
a^{\prime }\left( t\right) $ and $b^{\prime }\left( t\right) .$

We choose $\mu _{0}=-2b^{\prime }\left( t\right) $ and the required solution
of the system is%
\begin{equation}
\alpha _{0}=-\frac{b\left( t\right) }{4b^{\prime }\left( t\right) },\qquad
\beta _{0}=\frac{1}{2b^{\prime }\left( t\right) },\qquad \gamma _{0}=-\frac{%
a^{\prime }\left( t\right) }{4b^{\prime }\left( t\right) }.  \label{mom7}
\end{equation}%
The Green function is given by%
\begin{equation}
G\left( x,y,t\right) =\frac{1}{\sqrt{-4\pi ib^{\prime }\left( t\right) }}%
\exp \left( \frac{b\left( t\right) x^{2}-2xy+a^{\prime }\left( t\right) y^{2}%
}{4ib^{\prime }\left( t\right) }\right) ,\qquad t>0.  \label{mom8}
\end{equation}%
A more general particular solution has the form (\ref{dec13}), where $\mu
=c_{1}a^{\prime }\left( t\right) +c_{2}b^{\prime }\left( t\right) ,$ $\mu
\left( 0\right) =c_{1}\neq 0$ and%
\begin{eqnarray}
&&\alpha =-\frac{1}{4}\frac{c_{1}a\left( t\right) +c_{2}b\left( t\right) }{%
c_{1}a^{\prime }\left( t\right) +c_{2}b^{\prime }\left( t\right) },\qquad
\alpha \left( 0\right) =-\frac{c_{1}}{4c_{2}},  \label{mom9} \\
&&\beta =\frac{c_{1}\beta \left( 0\right) }{c_{1}a^{\prime }\left( t\right)
+c_{2}b^{\prime }\left( t\right) },  \label{mom10} \\
&&\gamma =\gamma \left( 0\right) +\frac{c_{1}\beta ^{2}\left( 0\right)
b^{\prime }\left( t\right) }{c_{1}a^{\prime }\left( t\right) +c_{2}b^{\prime
}\left( t\right) }.  \label{mom11}
\end{eqnarray}%
This can be verified, once again, by a direct substitution into the system (%
\ref{mom2})--(\ref{mom4}) or with the aid of the transformations (\ref{dec17}%
)--(\ref{dec20}) and (\ref{dec21})--(\ref{dec24}).

The oscillatory case is similar. The Schr\"{o}dinger equation (\ref{osc1})
in the momentum representation has the form%
\begin{equation}
i\frac{\partial \psi }{\partial t}+t\frac{\partial ^{2}\psi }{\partial x^{2}}%
-\frac{1}{4}x^{2}\psi =0  \label{mom12}
\end{equation}%
and%
\begin{eqnarray}
&&\frac{d\alpha }{dt}+\frac{1}{4}+4t\alpha ^{2}=0,  \label{mom13} \\
&&\frac{d\beta }{dt}+4t\alpha \beta =0,  \label{mom14} \\
&&\frac{d\gamma }{dt}+t\beta ^{2}=0.  \label{mom15}
\end{eqnarray}%
Here%
\begin{equation}
\alpha =\frac{1}{4t}\frac{\mu ^{\prime }}{\mu }  \label{mom16}
\end{equation}%
and%
\begin{equation}
\mu ^{\prime \prime }-\frac{1}{t}\mu ^{\prime }+t\mu =0.  \label{mom17}
\end{equation}%
The corresponding solutions are%
\begin{equation}
\mu _{0}=2b^{\prime }\left( -t\right) ,\quad \alpha _{0}=\frac{b\left(
-t\right) }{4b^{\prime }\left( -t\right) },\quad \beta _{0}=-\frac{1}{%
2b^{\prime }\left( -t\right) },\quad \gamma _{0}=\frac{a^{\prime }\left(
-t\right) }{4b^{\prime }\left( -t\right) }  \label{mom18}
\end{equation}%
and%
\begin{eqnarray}
&&\mu =c_{1}a^{\prime }\left( -t\right) +c_{2}b^{\prime }\left( -t\right)
,\qquad \mu \left( 0\right) =c_{1}\neq 0,  \label{mom19} \\
&&\alpha =\frac{1}{4}\frac{c_{1}a\left( -t\right) +c_{2}b\left( -t\right) }{%
c_{1}a^{\prime }\left( -t\right) +c_{2}b^{\prime }\left( -t\right) },\qquad
\alpha \left( 0\right) =\frac{c_{2}}{4c_{1}},  \label{mom20} \\
&&\beta =\frac{c_{1}\beta \left( 0\right) }{c_{1}a^{\prime }\left( -t\right)
+c_{2}b^{\prime }\left( -t\right) },  \label{mom21} \\
&&\gamma =\gamma \left( 0\right) -\frac{c_{1}\beta ^{2}\left( 0\right)
b^{\prime }\left( -t\right) }{c_{1}a^{\prime }\left( -t\right)
+c_{2}b^{\prime }\left( -t\right) }.  \label{mom22}
\end{eqnarray}%
The Green function is given by%
\begin{equation}
G\left( x,y,t\right) =\frac{1}{\sqrt{4\pi ib^{\prime }\left( -t\right) }}%
\exp \left( i\frac{b\left( -t\right) x^{2}-2xy+a^{\prime }\left( -t\right)
y^{2}}{4b^{\prime }\left( -t\right) }\right) ,\qquad t>0.  \label{mom23}
\end{equation}%
We leave further details to the reader.

\section{Gauge Transformation}

The time-dependent Schr\"{o}dinger equation%
\begin{equation}
i\frac{\partial \psi }{\partial t}=\left( \frac{1}{4}\left( p-A\left(
x,t\right) \right) ^{2}+V\left( x,t\right) \right) \psi ,  \label{g1}
\end{equation}%
where $p=i^{-1}\partial /\partial x$ is the linear momentum operator, with
the help of the gauge transformation%
\begin{equation}
\psi =e^{-if\left( x,t\right) }\psi ^{\prime }  \label{g2}
\end{equation}%
can be transformed into a similar form%
\begin{equation}
i\frac{\partial \psi ^{\prime }}{\partial t}=\left( \frac{1}{4}\left(
p-A^{\prime }\left( x,t\right) \right) ^{2}+V^{\prime }\left( x,t\right)
\right) \psi ^{\prime }  \label{g3}
\end{equation}%
with the new vector and scalar potentials given by%
\begin{equation}
A^{\prime }=A+\frac{\partial f}{\partial x},\qquad V^{\prime }=V-\frac{%
\partial f}{\partial t}.  \label{g4}
\end{equation}%
Here we consider the one-dimensional case only; see Refs.~\cite{La:Lif} and 
\cite{Merz} for more details.

An interesting special case of the gauge transformation related to this
paper is given by%
\begin{eqnarray}
&&A=0,\qquad V=-tx^{2},\qquad f=-\frac{x^{2}}{t}  \label{g5} \\
&&A^{\prime }=-\frac{2x}{t},\qquad V^{\prime }=-tx^{2}-\frac{x^{2}}{t^{2}},
\label{g6}
\end{eqnarray}%
when the new Hamiltonian is%
\begin{eqnarray}
H^{\prime } &=&\frac{1}{4}\left( p-A^{\prime }\right) ^{2}+V^{\prime }=\frac{%
1}{4}\left( p+\frac{2x}{t}\right) ^{2}-tx^{2}-\frac{x^{2}}{t^{2}}  \label{g7}
\\
&=&\frac{1}{4}\left( p^{2}+\frac{2}{t}\left( px+xp\right) +\frac{4x^{2}}{%
t^{2}}\right) -tx^{2}-\frac{x^{2}}{t^{2}}  \notag \\
&=&-\frac{1}{4}\frac{\partial ^{2}}{\partial x^{2}}-\frac{i}{2t}\left( 2x%
\frac{\partial }{\partial x}+1\right) -tx^{2},  \notag
\end{eqnarray}%
and equation (\ref{dec1}) takes the form%
\begin{equation}
i\frac{\partial \psi }{\partial t}+\frac{1}{4}\frac{\partial ^{2}\psi }{%
\partial x^{2}}+tx^{2}\psi +\frac{i}{2t}\left( 2x\frac{\partial \psi }{%
\partial x}+\psi \right) =0  \label{g8}
\end{equation}%
with a singular variable coefficient at the origin. Substitution (\ref{dec2}%
) results in%
\begin{eqnarray}
&&\frac{d\alpha }{dt}-t+\frac{2}{t}\alpha +\alpha ^{2}=0,  \label{g8a} \\
&&\frac{d\beta }{dt}+\left( \alpha +\frac{1}{t}\right) \beta =0,  \label{g8b}
\\
&&\frac{d\gamma }{dt}+\frac{1}{4}\beta ^{2}=0,  \label{g8c}
\end{eqnarray}%
where%
\begin{equation}
\alpha =\frac{\mu ^{\prime }}{\mu }-\frac{1}{t},\qquad \mu ^{\prime \prime
}-t\mu =0.  \label{g9d}
\end{equation}%
As a result one can conclude that the time-dependent Schr\"{o}dinger
equation (\ref{g8}) has a solution of the form%
\begin{equation}
\psi \left( x,t\right) =e^{-ix^{2}/t}\int_{-\infty }^{\infty }G\left(
x,y,t\right) \ \varphi \left( y\right) \ dy,  \label{g9}
\end{equation}%
where the Green function $G\left( x,y,t\right) $ is given by (\ref{dec11}).
This solution is not continuous when $t\rightarrow 0^{+}$ but it does
satisfy the following modified initial condition%
\begin{equation}
\lim_{t\rightarrow 0^{+}}e^{ix^{2}/t}\psi \left( x,t\right) =\varphi \left(
x\right) ,  \label{g10}
\end{equation}%
which reveals the structure of the singularity of the corresponding wave
function at the origin. We leave further details to the reader.

\section{Particular Solutions of Nonlinear Schr\"{o}dinger Equations}

One can find solutions of the corresponding nonlinear Schr\"{o}dinger
equations following Refs.~\cite{Cor-Sot:Lop:Sua:Sus} and \cite{Cor-Sot:Sus}.
For example, consider the case%
\begin{equation}
i\frac{\partial \psi }{\partial t}+\frac{1}{4}\frac{\partial ^{2}\psi }{%
\partial x^{2}}+tx^{2}\psi =h\left( t\right) \left\vert \psi \right\vert
^{2s}\psi ,\qquad s\geq 0  \label{ns1}
\end{equation}%
and look for a particular solution of the form%
\begin{equation}
\psi =\psi \left( x,t\right) =K_{h}\left( x,y,t\right) =\frac{e^{i\phi }}{%
\sqrt{\mu \left( t\right) }}\ e^{i\left( \alpha \left( t\right) x^{2}+\beta
\left( t\right) xy+\gamma \left( t\right) y^{2}+\kappa \left( t\right)
\right) },\qquad \phi =\text{constant}.  \label{ns2}
\end{equation}%
Then equations (\ref{dec3})--(\ref{dec5}) hold with the general solution
given by (\ref{dec14})--(\ref{dec16}). In addition,%
\begin{equation}
\frac{d\kappa }{dt}=-\frac{h\left( t\right) }{\mu ^{s}\left( t\right) }%
,\qquad \kappa \left( t\right) =\kappa \left( 0\right) -\int_{0}^{t}\frac{%
h\left( \tau \right) }{\mu ^{s}\left( \tau \right) }\ d\tau .  \label{ns3}
\end{equation}%
The last integral can be explicitly evaluated in some special cases, say,
when $h\left( t\right) =\lambda \mu ^{\prime }\left( t\right) :$%
\begin{equation}
\kappa \left( t\right) =\left\{ 
\begin{array}{ll}
\kappa \left( 0\right) -\dfrac{\lambda }{1-s}\left( \mu ^{1-s}\left(
t\right) -\mu ^{1-s}\left( 0\right) \right) , & \text{when }s\neq 1,\bigskip
\\ 
\kappa \left( 0\right) -\lambda \ln \left( \dfrac{\mu \left( t\right) }{\mu
\left( 0\right) }\right) , & \text{when }s=1.%
\end{array}%
\right.  \label{ns4}
\end{equation}%
Here $\mu \left( 0\right) \neq 0;$ see \cite{Cor-Sot:Lop:Sua:Sus} and \cite%
{Cor-Sot:Sus} for more details. An example of a discontinuity of the initial
data can be constructed by the method of Ref.~\cite{Cor-Sot:Sus}. Other
cases are investigated in a similar fashion.

\section{Quantum Parametric Oscillator and Airy Functions}

The time-dependent Schr\"{o}dinger equation for a parametric oscillator can
be written in the form%
\begin{equation}
i\hslash \frac{\partial \Psi }{\partial t}=H\Psi  \label{spc1}
\end{equation}%
with the Hamiltonian%
\begin{equation}
H=\frac{p^{2}}{2m}+\frac{m\omega ^{2}\left( t\right) }{2}x^{2},\qquad p=%
\frac{\hslash }{i}\frac{\partial }{\partial x},  \label{spc2}
\end{equation}%
where $\hslash $ is the Planck constant, $m$ is the mass of the particle, $%
\omega \left( t\right) $ is the time-dependent oscillation frequency. The
initial value problem of the form%
\begin{equation}
i\hslash \frac{\partial \Psi }{\partial t}=-\frac{\hslash ^{2}}{2m}\frac{%
\partial ^{2}\Psi }{\partial x^{2}}+\frac{m\omega ^{2}}{2}\left( \omega
t+\delta \right) x^{2}\ \Psi ,\qquad \Psi \left( x,0\right) =\Phi \left(
x\right)  \label{spc3}
\end{equation}%
can be solved by the technique from the previous sections in terms of Airy
functions. The substitution%
\begin{equation}
\Psi \left( x,t\right) =\varepsilon ^{1/2}\psi \left( \xi ,\tau \right)
\label{spc4}
\end{equation}%
with%
\begin{equation}
\tau =\omega t+\delta ,\qquad \xi =\varepsilon x,\qquad \varepsilon =\sqrt{%
\frac{m\omega }{2\hslash }}  \label{spc5}
\end{equation}%
results in%
\begin{equation}
i\frac{\partial \psi }{\partial \tau }+\frac{1}{4}\frac{\partial ^{2}\psi }{%
\partial \xi ^{2}}-\tau \xi ^{2}\psi =0,\qquad \psi \left( \xi ,\delta
\right) =\varphi \left( \xi \right) =\varepsilon ^{-1/2}\Phi \left( x\right)
.  \label{spc6}
\end{equation}%
The Green function has the form%
\begin{equation}
G\left( x,y,t\right) =\sqrt{\frac{m\omega }{4\pi i\hslash \mu \left( \tau
\right) }}\exp \left( i\dfrac{m\omega }{2\hslash }\left( \alpha \left( \tau
\right) x^{2}+\beta \left( \tau \right) xy+\gamma \left( \tau \right)
y^{2}\right) \right)  \label{spc7}
\end{equation}%
with $\tau =\omega t+\delta ,$ where $\mu \left( \delta \right) =0,$ $\mu
^{\prime }\left( \delta \right) =-\left( 1/2\right) W\left( ai,bi\right)
=1/2 $ and%
\begin{eqnarray}
\mu \left( \tau \right) &=&\frac{1}{2}\left( ai\left( -\delta \right)
bi\left( -\tau \right) -bi\left( -\delta \right) ai\left( -\tau \right)
\right) ,  \label{spc8} \\
\alpha \left( \tau \right) &=&-\frac{ai^{\prime }\left( -\tau \right)
bi\left( -\delta \right) -ai\left( -\delta \right) bi^{\prime }\left( -\tau
\right) }{ai\left( -\tau \right) bi\left( -\delta \right) -ai\left( -\delta
\right) bi\left( -\tau \right) },  \label{spc9} \\
\beta \left( \tau \right) &=&\frac{2}{ai\left( -\tau \right) bi\left(
-\delta \right) -ai\left( -\delta \right) bi\left( -\tau \right) },
\label{spc10} \\
\gamma \left( \tau \right) &=&-\frac{ai^{\prime }\left( -\delta \right)
bi\left( -\tau \right) -ai\left( -\tau \right) bi^{\prime }\left( -\delta
\right) }{ai\left( -\tau \right) bi\left( -\delta \right) -ai\left( -\delta
\right) bi\left( -\tau \right) }.  \label{spc11}
\end{eqnarray}%
This can be derived with the aid of transformation (\ref{dec21})--(\ref%
{dec24}). Thus%
\begin{equation}
\mu \sim \frac{1}{2}\omega t,\qquad \alpha \sim \frac{1}{\omega t},\qquad
\beta \sim -\frac{2}{\omega t},\qquad \gamma \sim \frac{1}{\omega t}
\label{spc10a}
\end{equation}%
as $t\rightarrow 0^{+}$ and the corresponding asymptotic formula is%
\begin{equation}
G\left( x,y,t\right) \sim \sqrt{\frac{m}{2\pi i\hslash t}}\exp \left( \frac{%
im\left( x-y\right) ^{2}}{2\hslash t}\right) ,\qquad t\rightarrow 0^{+},
\label{spc10b}
\end{equation}%
where expression on the right-hand side is a familiar free particle
propagator. The solution of the initial value problem (\ref{spc3}) is given
by%
\begin{equation}
\Psi \left( x,t\right) =\int_{-\infty }^{\infty }G\left( x,y,t\right) \ \Phi
\left( y\right) \ dy.  \label{spc11a}
\end{equation}%
We leave the calculation details to the reader and consider an application.

The time-dependent quadratic potential of the form%
\begin{equation}
V\left( x,t\right) =\left\{ 
\begin{array}{ll}
\dfrac{1}{2}m\omega _{0}^{2}\ x^{2}, & t\leq 0\medskip , \\ 
\dfrac{1}{2}m\omega ^{2}\left( \omega t+\delta \right) \ x^{2}, & 0\leq
t\leq T\medskip , \\ 
\dfrac{1}{2}m\omega _{1}^{2}\ x^{2}, & t\geq T\medskip 
\end{array}%
\right.   \label{spc12}
\end{equation}%
describes a parametric oscillator that changes its frequency from $\omega
_{0}$ to $\omega _{1}$ during the time interval $T.$ The continuity at $t=0$
and $t=T$ defines the transition parameters $\omega $ and $\delta $ as
follows%
\begin{equation}
\omega =\left( \frac{\omega _{1}^{2}-\omega _{0}^{2}}{T}\right)
^{1/3},\qquad \delta =\omega _{0}^{2}\left( \frac{T}{\omega _{1}^{2}-\omega
_{0}^{2}}\right) ^{2/3}  \label{spc13}
\end{equation}%
in terms of the initial $\omega _{0}$ and terminal $\omega _{1}$ oscillator
frequencies. This model is integrable in terms of Airy functions with the
help of the Green function found in this section as follows.

When $t<0$ the normalized wave function for a state with the definite energy 
$E_{n}^{\left( 0\right) }=\hslash \omega _{0}\left( n+1/2\right) $ is \cite%
{La:Lif}, \cite{Merz}:%
\begin{equation}
\Psi _{n}^{\left( 0\right) }\left( x,t\right) =\frac{e^{-i\omega _{0}\left(
n+1/2\right) t}}{\sqrt{2^{n}n!}}\left( \frac{m\omega _{0}}{\pi \hslash }%
\right) ^{1/4}\exp \left( -\frac{m\omega _{0}}{2\hslash }x^{2}\right) \
H_{n}\left( \sqrt{\frac{m\omega _{0}}{\hslash }}\ x\right) ,  \label{spc14}
\end{equation}%
where $H_{n}\left( \xi \right) $ are the Hermite polynomials \cite{Ab:St}, 
\cite{An:As:Ro}, \cite{Askey}, \cite{Ni:Su:Uv}, \cite{Ni:Uv}, \cite{Rain}, 
\cite{Sus:Trey}, and \cite{Sze}. When $0\leq t\leq T$ the corresponding
transition wave function is given by the time evolution operator%
\begin{equation}
\Psi _{n}\left( x,t\right) =U\left( t\right) \Psi _{n}^{\left( 0\right)
}=\int_{-\infty }^{\infty }G\left( x,y,t\right) \ \Psi _{n}^{\left( 0\right)
}\left( y,0\right) \ dy  \label{spc15}
\end{equation}%
with the Green function (\ref{spc7})--(\ref{spc11}). Finally, for $t\geq T$
the wave function is a linear combination%
\begin{equation}
\Psi _{n}\left( x,t\right) =\sum_{k=0}^{\infty }c_{kn}\left( T\right) \ \Psi
_{k}^{\left( 1\right) }\left( x,t\right)  \label{spc16}
\end{equation}%
of the eigenfunctions%
\begin{equation}
\Psi _{k}^{\left( 1\right) }\left( x,t\right) =\frac{e^{-i\omega _{1}\left(
k+1/2\right) \left( t-T\right) }}{\sqrt{2^{k}k!}}\left( \frac{m\omega _{1}}{%
\pi \hslash }\right) ^{1/4}\exp \left( -\frac{m\omega _{1}}{2\hslash }%
x^{2}\right) \ H_{k}\left( \sqrt{\frac{m\omega _{1}}{\hslash }}\ x\right)
\label{spc17}
\end{equation}%
corresponding to the new eigenvalues $E_{k}^{\left( 1\right) }=\hslash
\omega _{1}\left( k+1/2\right) $ with $k=0,1,2,...\ .$ Thus function $%
c_{kn}\left( T\right) $ gives the quantum mechanical amplitude that the
oscillator initially in state $\left( \omega _{0},n\right) $ is found at
time $T$ in state $\left( \omega _{1},k\right) .$

For the transition period $0\leq t\leq T$ use the integral%
\begin{equation}
\int_{-\infty }^{\infty }e^{-\lambda ^{2}\left( x-y\right) ^{2}}H_{n}\left(
ay\right) \ dy=\frac{\sqrt{\pi }}{\lambda ^{n+1}}\left( \lambda
^{2}-a^{2}\right) ^{n/2}H_{n}\left( \frac{\lambda ax}{\left( \lambda
^{2}-a^{2}\right) ^{1/2}}\right) ,\quad \func{Re}\lambda ^{2}>0,  \label{Erd}
\end{equation}%
which is equivalent to Eq.~(30) on page 195 of Vol.~2 of Ref.~\cite{Erd}
(the Gauss transform of Hermite polynomials), or Eq.~(17) on page 290 of
Vol.~2 of Ref.~\cite{ErdInt}. The initial wave function evolves in the
following manner%
\begin{eqnarray}
\Psi _{n}\left( x,t\right)  &=&i^{n}\left( \frac{m\omega _{0}}{\pi \hslash }%
\right) ^{1/4}\sqrt{\frac{\omega }{i\mu 2^{n+1}n!\left( \omega _{0}-i\gamma
\omega \right) }}\left( \frac{\omega _{0}+i\gamma \omega }{\omega
_{0}-i\gamma \omega }\right) ^{n/2}  \label{spc18} \\
&&\times \exp \left( i\frac{m\omega }{2\hslash }\left( \alpha -\frac{\omega
^{2}\beta ^{2}\gamma }{4\left( \omega _{0}^{2}+\gamma ^{2}\omega ^{2}\right) 
}\right) x^{2}\right)   \notag \\
&&\times \exp \left( -\frac{m\omega _{0}\omega ^{2}\beta ^{2}x^{2}}{8\hslash
\left( \omega _{0}^{2}+\gamma ^{2}\omega ^{2}\right) }\right) \ H_{n}\left( 
\sqrt{\frac{m\omega _{0}}{4\hslash \left( \omega _{0}^{2}+\gamma ^{2}\omega
^{2}\right) }}\ \omega \beta x\right) ,  \notag
\end{eqnarray}%
where the time-dependent coefficients $\mu ,$ $\alpha ,$ $\beta ,$ and $%
\gamma $ are given by equations (\ref{spc8})--(\ref{spc11}) in terms of Airy
functions with the argument $\tau =\omega t+\delta $ during the time
interval $0\leq t\leq T.$ The asymptotics (\ref{spc10a}) imply that $\Psi
_{n}\left( x,t\right) \rightarrow \Psi _{n}^{\left( 0\right) }\left(
x,0\right) $ as $t\rightarrow 0^{+}$ with the choice of principal branch of
the radicals. A direct integration shows that%
\begin{equation}
\int_{-\infty }^{\infty }\left\vert \Psi _{n}\left( x,t\right) \right\vert
^{2}\ dx=1,\qquad 0\leq t\leq T  \label{spc18a}
\end{equation}%
by the familiar orthogonality relation of the Hermite polynomials. The
normalization of the wave function holds also, of course, due to the
unitarity of the time evolution operator.

Then in view of the orthogonality of eigenfunctions (\ref{spc17}) the
transition amplitudes are%
\begin{equation}
c_{kn}\left( T\right) =\int_{-\infty }^{\infty }\left( \Psi _{k}^{\left(
1\right) }\left( x,T\right) \right) ^{\ast }\Psi _{n}\left( x,T\right) \ dx,
\label{spc19}
\end{equation}%
where one can use another classical integral evaluated by Bailey:%
\begin{eqnarray}
&&\int_{-\infty }^{\infty }e^{-\lambda ^{2}x^{2}}H_{m}\left( ax\right)
H_{n}\left( bx\right) \ dx  \label{Bai} \\
&&\quad =\frac{2^{m+n}}{\lambda ^{m+n+1}}\Gamma \left( \frac{m+n+1}{2}%
\right) \left( a^{2}-\lambda ^{2}\right) ^{m/2}\left( b^{2}-\lambda
^{2}\right) ^{n/2}  \notag \\
&&\qquad \times ~_{2}F_{1}\left( 
\begin{array}{c}
-m,\quad -n \\ 
\dfrac{1}{2}\left( 1-m-n\right)%
\end{array}%
;\dfrac{1}{2}\left( 1-\frac{ab}{\sqrt{\left( a^{2}-\lambda ^{2}\right)
\left( b^{2}-\lambda ^{2}\right) }}\right) \right) ,\quad \func{Re}\lambda
^{2}>0,  \notag
\end{eqnarray}%
if $m+n$ is even; the integral vanishes by symmetry if $m+n$ is odd; see
Refs.~\cite{Bailey48} and \cite{Lord49} and references therein for earlier
works on these integrals, their special cases and extensions.

The end result is $c_{kn}\left( T\right) =0,$ if $k+n$ is odd, and 
\begin{eqnarray}
c_{kn}\left( T\right) &=&i^{n}\Gamma \left( \frac{k+n+1}{2}\right) \left( 
\frac{\omega _{0}\omega _{1}}{\pi ^{2}}\right) ^{1/4}\sqrt{\frac{2^{k+n}}{%
i\mu k!n!\left( \omega _{0}-i\gamma \omega \right) }}\left( \frac{\omega
_{0}+i\gamma \omega }{\omega _{0}-i\gamma \omega }\right) ^{n/2}
\label{spc20} \\
&&\times \left( \frac{\omega _{1}}{\omega }-\frac{\omega _{0}\omega \beta
^{2}}{4\left( \omega _{0}^{2}+\gamma ^{2}\omega ^{2}\right) }+i\left( \alpha
-\frac{\omega ^{2}\beta ^{2}\gamma }{4\left( \omega _{0}^{2}+\gamma
^{2}\omega ^{2}\right) }\right) \right) ^{k/2}  \notag \\
&&\times \left( \frac{\omega _{0}\omega \beta ^{2}}{4\left( \omega
_{0}^{2}+\gamma ^{2}\omega ^{2}\right) }-\frac{\omega _{1}}{\omega }+i\left(
\alpha -\frac{\omega ^{2}\beta ^{2}\gamma }{4\left( \omega _{0}^{2}+\gamma
^{2}\omega ^{2}\right) }\right) \right) ^{n/2}  \notag \\
&&\times \left( \frac{\omega _{1}}{\omega }+\frac{\omega _{0}\omega \beta
^{2}}{4\left( \omega _{0}^{2}+\gamma ^{2}\omega ^{2}\right) }-i\left( \alpha
-\frac{\omega ^{2}\beta ^{2}\gamma }{4\left( \omega _{0}^{2}+\gamma
^{2}\omega ^{2}\right) }\right) \right) ^{-\left( k+n+1\right) /2}  \notag \\
&&\times ~_{2}F_{1}\left( 
\begin{array}{c}
-k,\quad -n \\ 
\dfrac{1}{2}\left( 1-k-n\right)%
\end{array}%
;\quad \frac{1}{2}\left( 1+i\zeta \right) \right) ,  \notag
\end{eqnarray}%
where%
\begin{equation}
\zeta =\frac{\omega \beta \sqrt{\omega _{0}\omega _{1}}}{\sqrt{\left( \alpha
\omega _{0}-\gamma \omega _{1}\right) ^{2}\omega ^{2}+\left( \omega
_{0}\omega _{1}+\alpha \gamma \omega ^{2}-\beta ^{2}\omega ^{2}/4\right) ^{2}%
}},  \label{spc21}
\end{equation}%
if $k+n$ is even. The terminating hypergeometric function is transformed as
follows%
\begin{eqnarray}
&&_{2}F_{1}\left( 
\begin{array}{c}
-k,\quad -n \\ 
\dfrac{1}{2}\left( 1-k-n\right)%
\end{array}%
;\quad \frac{1}{2}\left( 1+i\zeta \right) \right) \medskip  \label{spc21a} \\
&&\ =\left\{ 
\begin{array}{ll}
\dfrac{\left( 1/2\right) _{r}\left( 1/2\right) _{s}}{\left( 1/2\right) _{r+s}%
}\ _{2}F_{1}\left( 
\begin{array}{c}
-r,\quad -s\medskip \\ 
1/2%
\end{array}%
;\quad -\zeta ^{2}\right) ,\medskip & \text{if }k=2r,\ n=2s,\medskip \\ 
-\dfrac{\left( 3/2\right) _{r}\left( 3/2\right) _{s}}{\left( 3/2\right)
_{r+s}}\ i\zeta ~_{2}F_{1}\left( 
\begin{array}{c}
-r,\quad -s\medskip \\ 
3/2%
\end{array}%
;\quad -\zeta ^{2}\right) , & \text{if }k=2r+1,\ n=2s+1.%
\end{array}%
\right.  \notag
\end{eqnarray}%
It is valid in the entire complex plane; the details are given in
Appendix~B. The transformation (\ref{spc21a}) completes evaluation of the
Bailey integral (\ref{Bai}).

Our function $c_{kn}\left( T\right) $ gives explicitly the quantum
mechanical amplitude that the oscillator initially in state $\left( \omega
_{0},n\right) $ is found at time $T$ in state $\left( \omega _{1},k\right) .$
The unitarity of the time evolution operator implies the discrete
orthogonality relation%
\begin{equation}
\sum_{k=0}^{\infty }c_{kn}^{\ast }\left( T\right) \ c_{kp}\left( T\right)
=\delta _{np}  \label{spc22}
\end{equation}%
for $_{2}F_{1}$ functions under consideration. The well-known orthogonal
systems at this level are Jacobi, Kravchuk, Meixner and Meixner--Pollaczek
polynomials; see, for example, \cite{An:As}, \cite{An:As:Ro}, \cite{Askey}, 
\cite{AskeyCH}, \cite{As:WiCH}, \cite{As:Wi}, \cite{At:SusCH}, \cite{Erd}, 
\cite{Ko:Sw}, \cite{Ni:Su:Uv}, \cite{Ni:Uv}, \cite{SusCH}, \cite{Su1}, \cite%
{Sus:Trey}, \cite{Sze}, and references therein. This particular $_{2}F_{1}$
orthogonal system is reduced by the transformation (\ref{spc21a}) to the
Meixner polynomials. A group theoretical interpretation of the transition
amplitudes and their relation with Bargmann's functions is discussed in
section~10.

In the limit $T\rightarrow 0^{+},$ when the oscillator frequency changes
instantaneously from $\omega _{0}$ to $\omega _{1},$ the transition
amplitudes are essentially simplified. As a result $c_{kn}\left( 0\right)
=0, $ if $k+n$ is odd, and%
\begin{eqnarray}
c_{kn}\left( 0\right) &=&i^{n}\Gamma \left( \frac{k+n+1}{2}\right) \left( 
\frac{\omega _{0}\omega _{1}}{\pi ^{2}}\right) ^{1/4}\sqrt{\frac{2^{k+n+1}}{%
k!n!\left( \omega _{0}+\omega _{1}\right) }}\left( \frac{\omega _{1}-\omega
_{0}}{\omega _{1}+\omega _{0}}\right) ^{\left( k+n\right) /2}  \notag \\
&&\times ~_{2}F_{1}\left( 
\begin{array}{c}
-k,\quad -n \\ 
\dfrac{1}{2}\left( 1-k-n\right)%
\end{array}%
;\quad \frac{1}{2}\left( 1+2i\frac{\sqrt{\omega _{0}\omega _{1}}}{\left\vert
\omega _{0}-\omega _{1}\right\vert }\right) \right) ,  \label{spc23}
\end{eqnarray}%
if $k+n$ is even. The discrete orthogonality relation (\ref{spc22}) and
transformation (\ref{spc21a}) hold. The limit $\omega _{1}\rightarrow \omega
_{0}$ is interesting from the view point of perturbation theory.

If the oscillator is in the ground state $\left( \omega _{0},0\right) $
before the start of interaction, the transition probability of finding the
oscillator in the $n$th excited energy eigenstate $\left( \omega
_{1},n\right) $ with the new frequency is given by $\left\vert
c_{2k+1,0}\left( T\right) \right\vert ^{2}=0$ and%
\begin{eqnarray}
\left\vert c_{2k,0}\left( T\right) \right\vert ^{2} &=&\frac{\left\vert
\beta \right\vert \omega \sqrt{\omega _{0}\omega _{1}}}{\sqrt{\left( \alpha
\omega _{0}+\gamma \omega _{1}\right) ^{2}\omega ^{2}+\left( \omega
_{0}\omega _{1}-\alpha \gamma \omega ^{2}+\beta ^{2}\omega ^{2}/4\right) ^{2}%
}}\   \label{spc24} \\
&&\times \frac{\left( 1/2\right) _{k}}{k!}\left( \frac{\left( \alpha \omega
_{0}-\gamma \omega _{1}\right) ^{2}\omega ^{2}+\left( \omega _{0}\omega
_{1}+\alpha \gamma \omega ^{2}-\beta ^{2}\omega ^{2}/4\right) ^{2}}{\left(
\alpha \omega _{0}+\gamma \omega _{1}\right) ^{2}\omega ^{2}+\left( \omega
_{0}\omega _{1}-\alpha \gamma \omega ^{2}+\beta ^{2}\omega ^{2}/4\right) ^{2}%
}\right) ^{k},  \notag
\end{eqnarray}%
where $k=0,1,2,...\ $and $\sum_{k=0}^{\infty }\left\vert c_{2k,0}\left(
T\right) \right\vert ^{2}=1$ with the help of binomial theorem. If the
oscillator is in the first excited state $\left( \omega _{0},1\right) ,$ the
transition probability of finding the oscillator in the $n$th excited state $%
\left( \omega _{1},n\right) $ is given by $\left\vert c_{2k,1}\left(
T\right) \right\vert ^{2}=0$ and%
\begin{eqnarray}
\left\vert c_{2k+1,1}\left( T\right) \right\vert ^{2} &=&\left( \frac{\beta
^{2}\omega ^{2}\omega _{0}\omega _{1}}{\left( \alpha \omega _{0}+\gamma
\omega _{1}\right) ^{2}\omega ^{2}+\left( \omega _{0}\omega _{1}-\alpha
\gamma \omega ^{2}+\beta ^{2}\omega ^{2}/4\right) ^{2}}\right) ^{3/2}
\label{spc24a} \\
&&\times \frac{\left( 3/2\right) _{k}}{k!}\left( \frac{\left( \alpha \omega
_{0}-\gamma \omega _{1}\right) ^{2}\omega ^{2}+\left( \omega _{0}\omega
_{1}+\alpha \gamma \omega ^{2}-\beta ^{2}\omega ^{2}/4\right) ^{2}}{\left(
\alpha \omega _{0}+\gamma \omega _{1}\right) ^{2}\omega ^{2}+\left( \omega
_{0}\omega _{1}-\alpha \gamma \omega ^{2}+\beta ^{2}\omega ^{2}/4\right) ^{2}%
}\right) ^{k},  \notag
\end{eqnarray}%
where $k=0,1,2,...\ $and $\sum_{k=0}^{\infty }\left\vert c_{2k+1,1}\left(
T\right) \right\vert ^{2}=1.$ These probabilities can be recognized as two
special cases of the negative binomial distribution, or Pascal distribution,
which gives the normalized weight function for the Meixner polynomials of a
discrete variable \cite{An:As:Ro}, \cite{Askey},\cite{Erd}, \cite{Ni:Su:Uv}, 
\cite{Ni:Uv}, and \cite{Sze}.

In a similar fashion, the probability that the oscillator initially in
eigenstate $\left( \omega _{0},n\right) $ is found at time $T$ after the
transition in state $\left( \omega _{1},k\right) $ is given by $\left\vert
c_{kn}\left( T\right) \right\vert ^{2}=0,$ if $k+n$ is odd, and%
\begin{eqnarray}
\left\vert c_{kn}\left( T\right) \right\vert ^{2} &=&\frac{\left\vert \beta
\right\vert \omega \sqrt{\omega _{0}\omega _{1}}}{\sqrt{\left( \alpha \omega
_{0}+\gamma \omega _{1}\right) ^{2}\omega ^{2}+\left( \omega _{0}\omega
_{1}-\alpha \gamma \omega ^{2}+\beta ^{2}\omega ^{2}/4\right) ^{2}}}
\label{spc25} \\
&&\times \frac{2^{k+n}}{k!n!\pi }\Gamma ^{2}\left( \frac{k+n+1}{2}\right) 
\notag \\
&&\times \left( \frac{\left( \alpha \omega _{0}-\gamma \omega _{1}\right)
^{2}\omega ^{2}+\left( \omega _{0}\omega _{1}+\alpha \gamma \omega
^{2}-\beta ^{2}\omega ^{2}/4\right) ^{2}}{\left( \alpha \omega _{0}+\gamma
\omega _{1}\right) ^{2}\omega ^{2}+\left( \omega _{0}\omega _{1}-\alpha
\gamma \omega ^{2}+\beta ^{2}\omega ^{2}/4\right) ^{2}}\right) ^{\left(
k+n\right) /2}  \notag \\
&&\times \left\vert ~_{2}F_{1}\left( 
\begin{array}{c}
-k,\quad -n \\ 
\dfrac{1}{2}\left( 1-k-n\right)%
\end{array}%
;\quad \frac{1}{2}\left( 1+i\zeta \right) \right) \right\vert ^{2},  \notag
\end{eqnarray}%
if $k+n$ is even. The transformation (\ref{spc21a}) is, of course, valid but
the square of the hypergeometric function can be simplified to a single
positive sum with the help of the quadratic transformation (\ref{quad})
followed by the Clausen formula (\ref{claus}):%
\begin{eqnarray}
&&\left\vert ~_{2}F_{1}\left( 
\begin{array}{c}
-k,\quad -n \\ 
\dfrac{1}{2}\left( 1-k-n\right)%
\end{array}%
;\quad \frac{1}{2}\left( 1+i\zeta \right) \right) \right\vert ^{2}
\label{spc26} \\
&&\qquad =~_{3}F_{2}\left( 
\begin{array}{c}
-k,\quad -n,\quad -\left( k+n\right) /2\medskip \\ 
\left( 1-k-n\right) /2,\quad -k-n%
\end{array}%
;\quad z\right)  \notag
\end{eqnarray}%
with%
\begin{equation}
z=\frac{\left( \alpha \omega _{0}+\gamma \omega _{1}\right) ^{2}\omega
^{2}+\left( \omega _{0}\omega _{1}-\alpha \gamma \omega ^{2}+\beta
^{2}\omega ^{2}/4\right) ^{2}}{\left( \alpha \omega _{0}-\gamma \omega
_{1}\right) ^{2}\omega ^{2}+\left( \omega _{0}\omega _{1}+\alpha \gamma
\omega ^{2}-\beta ^{2}\omega ^{2}/4\right) ^{2}}.  \label{spc27}
\end{equation}%
More details are given in the next section.

Thus we have determined a complete dynamics of the quantum parametric
oscillator transition from the initial state with the frequency $\omega _{0}$
to the terminal one with the frequency $\omega _{1}$ by explicitly solving
the time-dependent Schr\"{o}dinger equation with variable potential (\ref%
{spc12}) at all times.

%
%
\begin{figure}[htbp]
\centering \scalebox{1.0}{\includegraphics{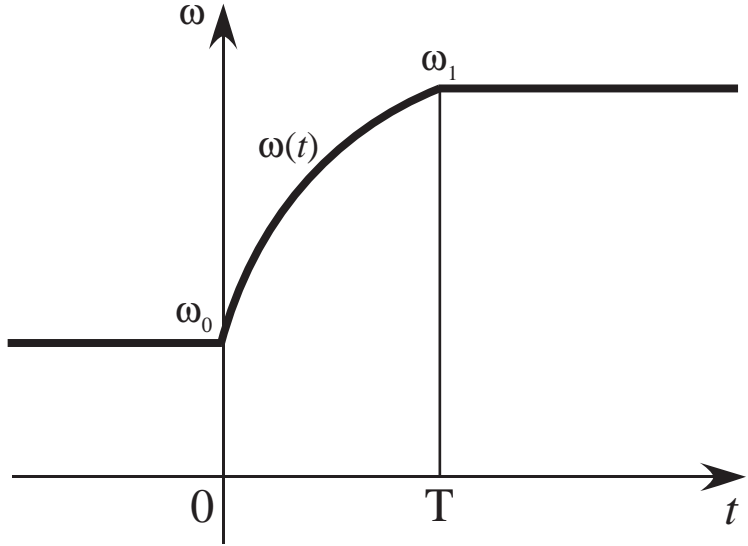}}
\caption{The parametric oscillator frequency.}
\end{figure}
%

\section{Quantum Parametric Oscillator: General Case}

The general case of the parametric oscillator with a variable frequency of
the form:%
\begin{equation}
\omega \left( t\right) =\left\{ 
\begin{array}{ll}
\omega _{0}, & t\leq 0\medskip , \\ 
\omega \left( t\right) , & 0\leq t\leq T\medskip , \\ 
\omega _{1}, & t\geq T\medskip%
\end{array}%
\right. \qquad \omega _{0}=\omega \left( 0\right) ,\quad \omega \left(
T\right) =\omega _{1}  \label{gen1}
\end{equation}%
(see Figure~1) can be investigated in a similar fashion. By the method of
Ref.~\cite{Cor-Sot:Lop:Sua:Sus} the corresponding transition Green function
is given by%
\begin{equation}
G\left( x,y,t\right) =\frac{1}{\sqrt{2\pi i\mu \left( t\right) }}\exp \left(
i\frac{m}{2\hslash }\left( \alpha \left( t\right) x^{2}+\beta \left(
t\right) xy+\gamma \left( t\right) y^{2}\right) \right) ,\qquad 0\leq t\leq
T,  \label{gen2}
\end{equation}%
where $\mu =\mu \left( t\right) $ is a solution of the equation of motion
for the classical parametric oscillator \cite{Lan:Lif}, \cite{Mag:Win}:%
\begin{equation}
\mu ^{\prime \prime }+\omega ^{2}\left( t\right) \mu =0,  \label{gen3}
\end{equation}%
that satisfies the initial conditions $\mu \left( 0\right) =0$ and $\mu
^{\prime }\left( 0\right) =$ $\hslash /m.$ The coefficients of the quadratic
form are%
\begin{eqnarray}
&&\alpha \left( t\right) =\frac{\mu ^{\prime }\left( t\right) }{\mu \left(
t\right) },\qquad \beta \left( t\right) =-\frac{2\hslash }{m\mu \left(
t\right) },  \label{gen4} \\
&&\gamma \left( t\right) =\frac{\hslash ^{2}}{m^{2}}\left( \frac{1}{\mu
\left( t\right) \mu ^{\prime }\left( t\right) }-\int_{0}^{t}\frac{\omega
^{2}\left( \tau \right) }{\left( \mu ^{\prime }\left( \tau \right) \right)
^{2}}\ d\tau \right) .  \label{gen5}
\end{eqnarray}%
The asymptotics (\ref{spc10b}) hold as $t\rightarrow 0^{+};$ see Ref.$~$\cite%
{Suaz:Sus} for more details.

A similar calculation gives the wave function (\ref{spc15}) during the
transition period $0<t\leq T$ as follows%
\begin{eqnarray}
\Psi _{n}\left( x,t\right) &=&\left( \dfrac{\hslash \omega _{0}}{\pi m}%
\right) ^{1/4}\frac{1}{\sqrt{\mu \ 2^{n}n!\left( \gamma +i\omega _{0}\right) 
}}\left( \frac{\gamma -i\omega _{0}}{\gamma +i\omega _{0}}\right) ^{n/2}
\label{gen6} \\
&&\times \exp \left( i\frac{m}{2\hslash }\left( \alpha -\frac{\beta
^{2}\gamma }{4\left( \gamma ^{2}+\omega _{0}^{2}\right) }\right) x^{2}\right)
\notag \\
&&\times \exp \left( -\frac{m\omega _{0}\beta ^{2}x^{2}}{8\hslash \left(
\gamma ^{2}+\omega _{0}^{2}\right) }\right) \ H_{n}\left( \sqrt{\dfrac{%
m\omega _{0}}{4\hslash \left( \gamma ^{2}+\omega _{0}^{2}\right) }}\ \beta
x\right) .  \notag
\end{eqnarray}%
It satisfies the normalization condition (\ref{spc18a}). The continuity
property $\Psi _{n}\left( x,t\right) \rightarrow \Psi _{n}^{\left( 0\right)
}\left( x,0\right) $ holds as $t\rightarrow 0^{+}$ for the principal branch
of the radicals.

The transition amplitudes (\ref{spc19}) are $c_{kn}\left( T\right) =0,$ if $%
k+n$ is odd, and%
\begin{eqnarray}
c_{kn}\left( T\right) &=&\Gamma \left( \frac{k+n+1}{2}\right) \left( \frac{%
\omega _{0}\omega _{1}}{\pi ^{2}}\right) ^{1/4}\sqrt{\frac{2^{k+n+1}\hslash 
}{\mu \ k!n!\left( \gamma +i\omega _{0}\right) m}}\left( \frac{\gamma
-i\omega _{0}}{\gamma +i\omega _{0}}\right) ^{n/2}  \label{gen7} \\
&&\times \left( \omega _{1}-\frac{\omega _{0}\beta ^{2}}{4\left( \gamma
^{2}+\omega _{0}^{2}\right) }+i\left( \alpha -\frac{\beta ^{2}\gamma }{%
4\left( \gamma ^{2}+\omega _{0}^{2}\right) }\right) \right) ^{k/2}  \notag \\
&&\times \left( -\omega _{1}+\frac{\omega _{0}\beta ^{2}}{4\left( \gamma
^{2}+\omega _{0}^{2}\right) }+i\left( \alpha -\frac{\beta ^{2}\gamma }{%
4\left( \gamma ^{2}+\omega _{0}^{2}\right) }\right) \right) ^{n/2}  \notag \\
&&\times \left( \omega _{1}+\frac{\omega _{0}\beta ^{2}}{4\left( \gamma
^{2}+\omega _{0}^{2}\right) }-i\left( \alpha -\frac{\beta ^{2}\gamma }{%
4\left( \gamma ^{2}+\omega _{0}^{2}\right) }\right) \right) ^{-\left(
k+n+1\right) /2}  \notag \\
&&\times ~_{2}F_{1}\left( 
\begin{array}{c}
-k,\quad -n \\ 
\dfrac{1}{2}\left( 1-k-n\right)%
\end{array}%
;\quad \frac{1}{2}\left( 1+i\zeta \right) \right) ,  \notag
\end{eqnarray}%
where%
\begin{equation}
\zeta =\frac{\beta \sqrt{\omega _{0}\omega _{1}}}{\sqrt{\left( \alpha \omega
_{0}-\gamma \omega _{1}\right) ^{2}+\left( \omega _{0}\omega _{1}+\alpha
\gamma -\beta ^{2}/4\right) ^{2}}},  \label{gen8}
\end{equation}%
if $k+n$ is even. The transformation (\ref{spc21a}) is applied and the
unitarity of the time evolution operator implies once again the discrete
orthogonality relation (\ref{spc22}) for the $_{2}F_{1}$ functions. Their
relations with Meixner polynomials and Bargmann's functions are discussed in
section~10.

For the oscillator initially in the ground state $\left( \omega
_{0},0\right) $ the transition probability of finding the oscillator in the $%
n$th excited energy eigenstate $\left( \omega _{1},n\right) $ with the new
frequency is given by $\left\vert c_{2k+1,0}\left( T\right) \right\vert
^{2}=0$ and%
\begin{eqnarray}
\left\vert c_{2k,0}\left( T\right) \right\vert ^{2} &=&\frac{\left\vert
\beta \right\vert \sqrt{\omega _{0}\omega _{1}}}{\sqrt{\left( \alpha \omega
_{0}+\gamma \omega _{1}\right) ^{2}+\left( \omega _{0}\omega _{1}-\alpha
\gamma +\beta ^{2}/4\right) ^{2}}}  \label{gen8aa} \\
&&\times \frac{\left( 1/2\right) _{k}}{k!}\left( \frac{\left( \alpha \omega
_{0}-\gamma \omega _{1}\right) ^{2}+\left( \omega _{0}\omega _{1}+\alpha
\gamma -\beta ^{2}/4\right) ^{2}}{\left( \alpha \omega _{0}+\gamma \omega
_{1}\right) ^{2}+\left( \omega _{0}\omega _{1}-\alpha \gamma +\beta
^{2}/4\right) ^{2}}\right) ^{k},  \notag
\end{eqnarray}%
where $k=0,1,2,...\ $and $\sum_{k=0}^{\infty }\left\vert c_{2k,0}\left(
T\right) \right\vert ^{2}=1.$ For the oscillator initially in the first
excited state $\left( \omega _{0},1\right) $ the transition probability of
finding the oscillator in the $n$th excited state $\left( \omega
_{1},n\right) $ is given by $\left\vert c_{2k,1}\left( T\right) \right\vert
^{2}=0$ and%
\begin{eqnarray}
\left\vert c_{2k+1,1}\left( T\right) \right\vert ^{2} &=&\left( \frac{\beta
^{2}\omega _{0}\omega _{1}}{\left( \alpha \omega _{0}+\gamma \omega
_{1}\right) ^{2}+\left( \omega _{0}\omega _{1}-\alpha \gamma +\beta
^{2}/4\right) ^{2}}\right) ^{3/2}  \label{gen8ab} \\
&&\times \frac{\left( 3/2\right) _{k}}{k!}\left( \frac{\left( \alpha \omega
_{0}-\gamma \omega _{1}\right) ^{2}+\left( \omega _{0}\omega _{1}+\alpha
\gamma -\beta ^{2}/4\right) ^{2}}{\left( \alpha \omega _{0}+\gamma \omega
_{1}\right) ^{2}+\left( \omega _{0}\omega _{1}-\alpha \gamma +\beta
^{2}/4\right) ^{2}}\right) ^{k},  \notag
\end{eqnarray}%
where $k=0,1,2,...\ $and $\sum_{k=0}^{\infty }\left\vert c_{2k,0}\left(
T\right) \right\vert ^{2}=1.$ Once again, these probabilities are two
special cases of the negative binomial distribution, which gives the
normalized weight function for the Meixner polynomials of a discrete
variable \cite{An:As:Ro}, \cite{Askey}, \cite{Erd}, \cite{Ni:Su:Uv}, \cite%
{Ni:Uv}, and \cite{Sze}.

In a similar fashion, the probability that the oscillator initially in
eigenstate $\left( \omega _{0},n\right) $ is found at time $T$ after the
transition in state $\left( \omega _{1},k\right) $ is given by $\left\vert
c_{kn}\left( T\right) \right\vert ^{2}=0,$ if $k+n$ is odd, and%
\begin{eqnarray}
\left\vert c_{kn}\left( T\right) \right\vert ^{2} &=&\frac{\left\vert \beta
\right\vert \sqrt{\omega _{0}\omega _{1}}}{\sqrt{\left( \alpha \omega
_{0}+\gamma \omega _{1}\right) ^{2}+\left( \omega _{0}\omega _{1}-\alpha
\gamma +\beta ^{2}/4\right) ^{2}}}  \label{gen9} \\
&&\times \frac{2^{k+n}}{k!n!\pi }\Gamma ^{2}\left( \frac{k+n+1}{2}\right) 
\notag \\
&&\times \left( \frac{\left( \alpha \omega _{0}-\gamma \omega _{1}\right)
^{2}+\left( \omega _{0}\omega _{1}+\alpha \gamma -\beta ^{2}/4\right) ^{2}}{%
\left( \alpha \omega _{0}+\gamma \omega _{1}\right) ^{2}+\left( \omega
_{0}\omega _{1}-\alpha \gamma +\beta ^{2}/4\right) ^{2}}\right) ^{\left(
k+n\right) /2}  \notag \\
&&\times \left\vert ~_{2}F_{1}\left( 
\begin{array}{c}
-k,\quad -n \\ 
\dfrac{1}{2}\left( 1-k-n\right)%
\end{array}%
;\quad \frac{1}{2}\left( 1+i\zeta \right) \right) \right\vert ^{2},  \notag
\end{eqnarray}%
if $k+n$ is even. The transformation (\ref{spc21a}) is valid once again but
the square of the hypergeometric function can be simplified to a single
positive sum with the help of a quadratic transformation (\ref{quad})
followed by the Clausen formula (\ref{claus}):%
\begin{eqnarray}
&&\left\vert ~_{2}F_{1}\left( 
\begin{array}{c}
-k,\quad -n \\ 
\dfrac{1}{2}\left( 1-k-n\right)%
\end{array}%
;\quad \frac{1}{2}\left( 1+i\zeta \right) \right) \right\vert ^{2}
\label{clausen} \\
&&\qquad =\left( _{2}F_{1}\left( 
\begin{array}{c}
-k/2,\quad -n/2\medskip \\ 
\left( 1-k-n\right) /2%
\end{array}%
;\quad 1+\zeta ^{2}\right) \right) ^{2}  \notag \\
&&\qquad =~_{3}F_{2}\left( 
\begin{array}{c}
-k,\quad -n,\quad -\left( k+n\right) /2\medskip \\ 
\left( 1-k-n\right) /2,\quad -k-n%
\end{array}%
;\quad z\right) ,  \notag
\end{eqnarray}%
where%
\begin{equation}
z=1+\zeta ^{2}=\frac{\left( \alpha \omega _{0}+\gamma \omega _{1}\right)
^{2}+\left( \omega _{0}\omega _{1}-\alpha \gamma +\beta ^{2}/4\right) ^{2}}{%
\left( \alpha \omega _{0}-\gamma \omega _{1}\right) ^{2}+\left( \omega
_{0}\omega _{1}+\alpha \gamma -\beta ^{2}/4\right) ^{2}}.  \label{argclaus}
\end{equation}%
Thus substituting (\ref{clausen})--(\ref{argclaus}) into (\ref{gen9}) one
gets the final representation of the probability $\left\vert c_{kn}\left(
T\right) \right\vert ^{2}$ in terms of a positive terminating $~_{3}F_{2}$
generalized hypergeometric function.

For an arbitrary initial data in $L^{2}\left( \boldsymbol{R}\right) :$%
\begin{equation}
\Psi ^{\left( 0\right) }\left( x,0\right) =\sum_{n=0}^{\infty }c_{n}^{\left(
0\right) }\ \Psi _{n}^{\left( 0\right) }\left( x,0\right) ,\qquad
\sum_{n=0}^{\infty }\left\vert c_{n}^{\left( 0\right) }\right\vert ^{2}=1,
\label{gen10}
\end{equation}%
the wave function after the transition is given by%
\begin{eqnarray}
\Psi ^{\left( 1\right) }\left( x,T\right) &=&\int_{-\infty }^{\infty
}G\left( x,y,T\right) \ \Psi ^{\left( 0\right) }\left( y,0\right) \ dy
\label{gen11} \\
&=&\sum_{n=0}^{\infty }c_{n}^{\left( 0\right) }\int_{-\infty }^{\infty
}G\left( x,y,T\right) \ \Psi _{n}^{\left( 0\right) }\left( y,0\right) \ dy 
\notag \\
&=&\sum_{n=0}^{\infty }c_{n}^{\left( 0\right) }\Psi _{n}\left( y,T\right)
=\sum_{n=0}^{\infty }c_{k}^{\left( 1\right) }\Psi _{k}^{\left( 1\right)
}\left( x,T\right)  \notag
\end{eqnarray}%
with%
\begin{equation}
c_{k}^{\left( 1\right) }=\sum_{n=0}^{\infty }c_{kn}\left( T\right) \
c_{n}^{\left( 0\right) }  \label{gen12}
\end{equation}%
by (\ref{spc16}). A group theoretical interpretation will be given in the
next section. The orthogonality property of the transition amplitudes (\ref%
{spc22}) implies the conservation law of the total probability%
\begin{equation}
\sum_{k=0}^{\infty }\left\vert c_{k}^{\left( 1\right) }\right\vert
^{2}=\sum_{n=0}^{\infty }\left\vert c_{n}^{\left( 0\right) }\right\vert
^{2}=1,  \label{gen12a}
\end{equation}%
which follows, of course, from the conservation of the norm of the wave
function during the transition.

Thus we have solved the problem of parametric oscillator in nonrelativistic
quantum mechanics provided that the solution of the corresponding classical
problem (\ref{gen3}) is known. A more convenient form of the transition
amplitudes (\ref{gen7}) will be given in the next section in terms of
Bargmann's functions. Moreover, the quantum forced parametric oscillator can
be investigated by the methods of Refs.~\cite{Cor-Sot:Lop:Sua:Sus}, \cite%
{Lop:Sus} and \cite{Me:Co:Su}. We leave the details to the reader.

\section{Group Theoretical Meaning of Transition Amplitudes}

The group theoretical properties of the harmonic oscillator wave functions
are investigated in detail. In addition to the well-known relation with the
Heisenberg--Weyl algebra of the creation and annihilation operators, the $n$%
-dimensional oscillator wave functions form a basis of the irreducible
unitary representation of the Lie algebra of the noncompact group $SU\left(
1,1\right) $ corresponding to the discrete positive series $\mathcal{D}%
_{+}^{j};$ see \cite{Fil:Ovch:Smir}, \cite{Me:Co:Su}, \cite{Ni:Su:Uv} and 
\cite{Smir:Shit}.\smallskip\ In this paper we are dealing with the
one-dimensional case only.

Define the creation and annihilation operators:%
\begin{equation}
a^{\dagger }\left( \omega \right) =\sqrt{\frac{m}{2\hslash \omega }}\left(
\omega x-\frac{\hslash }{m}\frac{\partial }{\partial x}\right) ,\qquad
a\left( \omega \right) =\sqrt{\frac{m}{2\hslash \omega }}\left( \omega x+%
\frac{\hslash }{m}\frac{\partial }{\partial x}\right) ,  \label{creann}
\end{equation}%
respectively, with the commutator:%
\begin{equation}
a\left( \omega \right) a^{\dagger }\left( \omega \right) -a^{\dagger }\left(
\omega \right) a\left( \omega \right) =1  \label{commut}
\end{equation}%
and the Hamiltonian:%
\begin{equation}
H\left( \omega \right) =\frac{\hslash \omega }{2}\left( a\left( \omega
\right) a^{\dagger }\left( \omega \right) +a^{\dagger }\left( \omega \right)
a\left( \omega \right) \right) .  \label{ham}
\end{equation}%
Their actions on \textquotedblleft stationary\textquotedblright\ oscillator
wave functions,%
\begin{equation}
\Psi _{n}\left( x\right) =\left( \frac{m\omega }{\pi \hslash }\right) ^{1/4}%
\frac{1}{\sqrt{2^{n}n!}}\exp \left( -\frac{m\omega }{2\hslash }x^{2}\right)
\ H_{n}\left( \sqrt{\frac{m\omega }{\hslash }}\ x\right) ,  \label{instwave}
\end{equation}%
are given by%
\begin{equation}
a\Psi _{n}=\sqrt{n}\ \Psi _{n-1},\qquad a^{\dagger }\Psi _{n}=\sqrt{n+1}\
\Psi _{n+1}.  \label{act}
\end{equation}%
Introducing operators%
\begin{align}
J_{+}\left( \omega \right) & =\frac{1}{2}\left( a^{\dagger }\left( \omega
\right) \right) ^{2},\qquad J_{-}\left( \omega \right) =\frac{1}{2}\left(
a\left( \omega \right) \right) ^{2},  \label{su1} \\
J_{0}\left( \omega \right) & =\frac{1}{4}\left( a\left( \omega \right)
a^{\dagger }\left( \omega \right) +a^{\dagger }\left( \omega \right) a\left(
\omega \right) \right) =\frac{1}{2\hslash \omega }H\left( \omega \right) ,
\label{su0}
\end{align}%
one can easily verify the following commutation relations:%
\begin{equation}
\left[ J_{0},J_{\pm }\right] =\pm J_{\pm },\qquad \left[ J_{+},J_{-}\right]
=-2J_{0}.  \label{comma}
\end{equation}%
For the Hermitian operators%
\begin{equation}
J_{x}=\frac{1}{2}\left( J_{+}+J_{-}\right) ,\qquad J_{y}=\frac{1}{2i}\left(
J_{+}-J_{-}\right) ,\qquad J_{z}=J_{0},  \label{commb}
\end{equation}%
we get%
\begin{equation}
\left[ J_{x},J_{y}\right] =-iJ_{z},\qquad \left[ J_{y},J_{z}\right]
=iJ_{x},\qquad \left[ J_{z},J_{x}\right] =iJ_{y}.  \label{commc}
\end{equation}%
These commutation rules are valid for the infinitesimal operators of the
non-compact group $SU\left( 1,1\right) ;$ see, for example, \cite{Bargmann47}%
, \cite{Fil:Ovch:Smir}, \cite{Me:Co:Su}, \cite{Ni:Su:Uv}, \cite{Smir:Shit},
and \cite{Vil} for more details.\smallskip

One can use a different notation for the oscillator wave functions (\ref%
{instwave}) as follows:%
\begin{equation}
\psi _{j\lambda }=\left\{ 
\begin{array}{ll}
\Psi _{2n}\left( x\right) ,\medskip  & \text{if }j=-3/4\text{ and }\lambda
=n+1/4, \\ 
\Psi _{2n+1}\left( x\right) , & \text{if }j=-1/4\text{ and }\lambda =n+3/4,%
\end{array}%
\right.   \label{not}
\end{equation}%
where $n=0,1,2,\ ...\ $and the inequality $\lambda \geq j+1$
holds.\smallskip\ The operators $J_{\pm }$ and $J_{0}$ have an explicit form%
\begin{equation}
J_{\pm }=\frac{1}{2}\left( \frac{m\omega }{\hslash }x^{2}-\frac{H}{\hslash
\omega }\mp \frac{1}{2}\mp x\frac{\partial }{\partial x}\right) ,\qquad
J_{0}=\frac{H}{2\hslash \omega }  \label{expl}
\end{equation}%
and their actions on the oscillator wave functions are given by%
\begin{equation}
J_{\pm }\psi _{j\lambda }=\sqrt{\left( \lambda \mp j\right) \left( \lambda
\pm j\pm 1\right) }\ \psi _{j,\lambda \pm 1},\qquad J_{0}\psi _{j\lambda
}=\lambda \psi _{j\lambda },  \label{acti}
\end{equation}%
whence%
\begin{equation}
J^{2}\psi _{j\lambda }=j\left( j+1\right) \psi _{j\lambda }  \label{mom}
\end{equation}%
with $J^{2}=J_{0}^{2}+J_{0}-J_{-}J_{+}=J_{0}^{2}-J_{0}-J_{+}J_{-}\ .$ These
relations coincide with the formulas that define the action of the
infinitesimal operators $J_{\pm }$ and $J_{0}$ of the Lie group $SU\left(
1,1\right) $ on a basis $\left\vert j,\lambda \right\rangle $ of the
irreducible representation $\mathcal{D}_{+}^{j}$ belonging to the discrete
positive series in an abstract Hilbert space \cite{Bargmann47}. Thus the
even and odd wave functions $\Psi _{n}\left( x\right) $ of the
one-dimensional harmonic oscillator form, respectively, bases for the two
irreducible representations $\mathcal{D}_{+}^{j}$ of the algebra $SU\left(
1,1\right) $ with the moments $j=-3/4$ for the even values of $n$ and $j=-1/4
$ for odd $n.$ It corresponds to the double valued representations of the
group $SU\left( 1,1\right) ,$ or quadruple valued representations of the $%
SO\left( 2,1\right) ;$ see \cite{Ni:Su:Uv} and \cite{Smir:Shit} for more
details. It is important for the purpose of this paper that these group
theoretical properties are valid \textquotedblleft
instantaneously\textquotedblright , when $\omega =\omega \left( t\right) $
is an arbitrary function of time.\ 

As a result of an elementary, but rather tedious, calculation our transition
amplitudes (\ref{gen7}) can be rewritten in the form%
\begin{equation}
c_{kn}\left( T\right) =T_{\lambda \lambda ^{\prime }}^{j}\left( \theta ,\tau
,\varphi \right)  \label{group1}
\end{equation}%
with the new $SU\left( 1,1\right) $ quantum numbers%
\begin{equation}
j=-\frac{3}{4},\qquad \lambda =r+\frac{1}{4},\qquad \lambda ^{\prime }=s+%
\frac{1}{4},  \label{group2}
\end{equation}%
if $k=2r,$ $n=2s,$ and%
\begin{equation}
j=-\frac{1}{4},\qquad \lambda =r+\frac{3}{4},\qquad \lambda ^{\prime }=s+%
\frac{3}{4},  \label{group3}
\end{equation}%
if $k=2r+1,$ $n=2s+1.$ The matrix elements $T_{\lambda \lambda ^{\prime
}}^{j}\left( \theta ,\tau ,\varphi \right) $ are the so-called Bargmann
functions, or the generalized spherical harmonics of $SU\left( 1,1\right) $ 
\cite{Bargmann47}, \cite{Ni:Su:Uv} and \cite{Vil}:%
\begin{equation}
T_{\lambda \lambda ^{\prime }}^{j}\left( \theta ,\tau ,\varphi \right)
=e^{-i\lambda \theta }t_{\lambda \lambda ^{\prime }}^{j}\left( \tau \right)
e^{-i\lambda ^{\prime }\varphi }.  \label{bargm1}
\end{equation}%
Here%
\begin{align}
& t_{\lambda \lambda ^{\prime }}^{j}\left( \tau \right) =\frac{\left(
-1\right) ^{\lambda -j-1}}{\Gamma \left( 2j+2\right) }\sqrt{\frac{\Gamma
\left( \lambda +j+1\right) \Gamma \left( \lambda ^{\prime }+j+1\right) }{%
\left( \lambda -j-1\right) !\left( \lambda ^{\prime }-j-1\right) !}}\ \left(
\sinh \frac{\tau }{2}\right) ^{-2j-2}\left( \tanh \frac{\tau }{2}\right)
^{\lambda +\lambda ^{\prime }}  \label{bargm2} \\
& \ \ \ \ \qquad \qquad \times \ \ _{2}F_{1}\left( 
\begin{array}{c}
-\lambda +j+1,\ -\lambda ^{\prime }+j+1\medskip \\ 
2j+2%
\end{array}%
;\ -\frac{1}{\sinh ^{2}\left( \tau /2\right) }\right)  \notag
\end{align}%
and the corresponding angles are given by%
\begin{eqnarray}
\tan \theta &=&\frac{2\alpha \omega _{0}^{2}\omega _{1}+2\gamma \omega
_{1}\left( \alpha \gamma -\beta ^{2}/4\right) }{\left( \alpha \omega
_{0}+\gamma \omega _{1}\right) \left( \alpha \omega _{0}-\gamma \omega
_{1}\right) -\left( \omega _{0}\omega _{1}+\alpha \gamma -\beta
^{2}/4\right) \left( \omega _{0}\omega _{1}-\alpha \gamma +\beta
^{2}/4\right) },  \label{bargm3} \\
\tan \varphi &=&\frac{-2\gamma \omega _{0}\omega _{1}^{2}-2\alpha \omega
_{0}\left( \alpha \gamma -\beta ^{2}/4\right) }{\left( \alpha \omega
_{0}+\gamma \omega _{1}\right) \left( \alpha \omega _{0}-\gamma \omega
_{1}\right) +\left( \omega _{0}\omega _{1}+\alpha \gamma -\beta
^{2}/4\right) \left( \omega _{0}\omega _{1}-\alpha \gamma +\beta
^{2}/4\right) }  \label{bargm4}
\end{eqnarray}%
for two Euclidean rotations, and%
\begin{equation}
\tanh ^{2}\left( \frac{\tau }{2}\right) =\frac{\left( \alpha \omega
_{0}-\gamma \omega _{1}\right) ^{2}+\left( \omega _{0}\omega _{1}+\alpha
\gamma -\beta ^{2}/4\right) ^{2}}{\left( \alpha \omega _{0}+\gamma \omega
_{1}\right) ^{2}+\left( \omega _{0}\omega _{1}-\alpha \gamma +\beta
^{2}/4\right) ^{2}}  \label{bargm5}
\end{equation}%
for the hyperbolic one. The branch of the radical is taken such that%
\begin{equation}
\frac{1}{\sinh \left( \tau /2\right) }=\frac{\beta \omega _{0}\omega _{1}}{%
\sqrt{\left( \alpha \omega _{0}-\gamma \omega _{1}\right) ^{2}+\left( \omega
_{0}\omega _{1}+\alpha \gamma -\beta ^{2}/4\right) ^{2}}}.  \label{bargm6}
\end{equation}%
The following symmetry holds: if $\alpha \leftrightarrow \gamma $ and $%
\omega _{0}\leftrightarrow \omega _{1},$ then $\theta \leftrightarrow
\varphi .$ It interchanges the initial and terminal oscillator states.

Our formulas (\ref{group1})--(\ref{bargm6}) give a clear group theoretical
interpretation of the transition amplitudes for the parametric oscillator in
quantum mechanics:%
\begin{eqnarray}
c_{kn}\left( T\right) &=&\left\langle \Psi _{k}^{\left( 1\right) }\left\vert
U\left( T\right) \Psi _{n}^{\left( 0\right) }\right. \right\rangle
\label{amp} \\
&=&\left\langle \psi _{j\lambda }^{\left( 1\right) }\left\vert T\left(
\theta ,\tau ,\varphi \right) \psi _{j\lambda ^{\prime }}^{\left( 0\right)
}\right. \right\rangle  \notag \\
&=&\left\langle \psi _{j\lambda }^{\left( 1\right) }\left\vert e^{-i\theta
J_{0}\left( \omega _{1}\right) }e^{-i\left( \tau +\ln \left( \omega
_{1}/\omega _{0}\right) \right) J_{y}}e^{-i\varphi J_{0}\left( \omega
_{0}\right) }\psi _{j\lambda ^{\prime }}^{\left( 0\right) }\right.
\right\rangle  \notag \\
&=&e^{-i\lambda \theta }\left\langle \psi _{j\lambda }^{\left( 1\right)
}\left\vert e^{-i\left( \tau +\ln \left( \omega _{1}/\omega _{0}\right)
\right) J_{y}}\psi _{j\lambda ^{\prime }}^{\left( 0\right) }\right.
\right\rangle e^{-i\lambda ^{\prime }\varphi }  \notag \\
&=&T_{\lambda \lambda ^{\prime }}^{j}\left( \theta ,\tau ,\varphi \right) , 
\notag
\end{eqnarray}%
if $k+n$ is even, in terms of the generalized spherical harmonics of the $%
SU\left( 1,1\right) $ algebra for the discrete positive series $\mathcal{D}%
_{+}^{j}$ with $j=-3/4$ and $j=-1/4$ for the even and odd oscillator
functions respectively. Formula (\ref{gen12}) gives a transformation of the
\textquotedblleft coordinates\textquotedblright\ of a wave function from the
old basis to the new one.

The time evolution operator has a familiar form%
\begin{equation}
U\left( t\right) =T\left( \theta ,\tau ,\varphi \right) =e^{-i\theta
J_{0}\left( \omega \right) }e^{-i\left( \tau +\ln \left( \omega /\omega
_{0}\right) \right) J_{y}}e^{-i\varphi J_{0}\left( \omega _{0}\right) }
\label{bargm7}
\end{equation}%
in terms of the corresponding infinitesimal operators, where%
\begin{equation}
-iJ_{y}=\frac{1}{4}+\frac{1}{2}x\frac{\partial }{\partial x}.  \label{bargm8}
\end{equation}%
The action of the hyperbolic rotation operator, or Lorentz boost, on a wave
function is given by%
\begin{equation}
e^{-i\tau J_{y}}\psi \left( x\right) =e^{\tau /4}\psi \left( e^{\tau
/2}x\right) .  \label{bargm9}
\end{equation}%
Therefore,%
\begin{eqnarray}
t_{\lambda \lambda ^{\prime }}^{j}\left( \tau \right) &=&\left\langle \psi
_{j\lambda }^{\left( 1\right) }\left\vert e^{-i\tau J_{y}}\psi _{j\lambda
^{\prime }}^{\left( 1\right) }\right. \right\rangle  \label{bargmint} \\
&=&e^{\tau /4}\int_{-\infty }^{\infty }\Psi _{k}^{\left( 1\right) }\left(
x\right) \ \Psi _{n}^{\left( 1\right) }\left( e^{\tau /2}x\right) \ dx, 
\notag
\end{eqnarray}%
if $k+n$ is even, which gives an integral representation for the Bargmann
function under consideration; cf.~\cite{Ni:Su:Uv}.

Thus the $SU\left( 1,1\right) $ symmetry suggests the following algebraic
form%
\begin{eqnarray}
U\left( t\right) &=&\exp \left( \frac{i\theta }{2\hslash \omega }\left( 
\frac{\hslash ^{2}}{2m}\frac{\partial ^{2}}{\partial x^{2}}-\frac{m\omega
^{2}}{2}x^{2}\right) \right)  \label{bargm10} \\
&&\times \exp \left( \frac{\tau +\ln \left( \omega /\omega _{0}\right) }{2}%
\left( \frac{1}{2}+x\frac{\partial }{\partial x}\right) \right)  \notag \\
&&\times \exp \left( \frac{i\varphi }{2\hslash \omega _{0}}\left( \frac{%
\hslash ^{2}}{2m}\frac{\partial ^{2}}{\partial x^{2}}-\frac{m\omega _{0}^{2}%
}{2}x^{2}\right) \right)  \notag
\end{eqnarray}%
of the time evolution operator, where the group parameters $\theta ,$ $\tau
, $ $\varphi $ are governed by the oscillator transition dynamics through
formulas (\ref{bargm3})--(\ref{bargm6}) back to the classical equation of
motion for the parametric oscillator (\ref{gen3}).

It is worth noting that Bargmann's functions are studied in detail. The
functions $t_{\lambda \lambda ^{\prime }}^{j}\left( \tau \right) $ are
related to the Meixner polynomials --- the unitarity property of the
Bargmann functions,%
\begin{equation}
\sum_{\lambda ^{\prime \prime }=j+1}^{\infty }t_{\lambda \lambda ^{\prime
\prime }}^{j}\left( \tau \right) t_{\lambda ^{\prime }\lambda ^{\prime
\prime }}^{j}\left( \tau \right) =\delta _{\lambda \lambda ^{\prime }},
\label{bargm11}
\end{equation}%
gives the discrete orthogonality relation of these polynomials. A connection
with a finite set of Jacobi polynomials orthogonal on an infinite interval
is also relevant. All basic facts about the functions $T_{\lambda \lambda
^{\prime }}^{j}\left( \theta ,\tau ,\varphi \right) $ can be derived from
the well-known properties of the Meixner and Jacobi polynomials; see Refs.~%
\cite{Ni:Su:Uv}, \cite{Smir:Sus:Shir}, and \cite{Vil} for more details.

\section{Summary}

The time-dependent Schr\"{o}dinger equations with variable coefficients%
\begin{equation}
i\frac{\partial \psi }{\partial t}+\frac{1}{4}\frac{\partial ^{2}\psi }{%
\partial x^{2}}\pm tx^{2}\psi =0  \label{sum1}
\end{equation}%
have the Green functions of the form%
\begin{equation}
G\left( x,y,t\right) =\frac{1}{\sqrt{\pm \pi ia\left( \pm t\right) }}\exp
\left( \pm i\frac{a^{\prime }\left( \pm t\right) -2xy+b\left( \pm t\right)
y^{2}}{a\left( \pm t\right) }\right) ,\qquad t>0,  \label{sum2}
\end{equation}%
where $a\left( t\right) =ai\left( t\right) $ and $b\left( t\right) =bi\left(
t\right) $ are solutions of the Airy equation $\mu ^{\prime \prime }-t\mu =0$
that satisfy the initial conditions $a\left( 0\right) =b^{\prime }\left(
0\right) =0$ and $a^{\prime }\left( 0\right) =b\left( 0\right) =1;$ see
Appendix~A below for construction of these solutions.

In the momentum representation the corresponding Schr\"{o}dinger equations
with variable coefficients%
\begin{equation}
i\frac{\partial \psi }{\partial t}\mp t\frac{\partial ^{2}\psi }{\partial
x^{2}}-\frac{1}{4}x^{2}\psi =0  \label{sum1a}
\end{equation}%
have the Green functions of the form%
\begin{equation}
G\left( x,y,t\right) =\frac{1}{\sqrt{\mp 4\pi ib^{\prime }\left( \pm
t\right) }}\exp \left( \mp i\frac{b\left( \pm t\right) -2xy+a^{\prime
}\left( \pm t\right) y^{2}}{4b^{\prime }\left( \pm t\right) }\right) ,\qquad
t>0,  \label{sum2a}
\end{equation}%
where $a^{\prime }\left( t\right) =ai^{\prime }\left( t\right) $ and $%
b^{\prime }\left( t\right) =bi^{\prime }\left( t\right) $ are solutions of
the equation $\mu ^{\prime \prime }-\left( 1/t\right) \mu ^{\prime }-t\mu =0$
that satisfy the initial conditions $a^{\prime }\left( 0\right) =1$ and $%
b^{\prime }\left( 0\right) =0;$ see Appendix~A for further properties of
these functions.

Solution of the corresponding Cauchy initial value problem is given by the
time evolution operator as follows%
\begin{equation}
\psi \left( x,t\right) =\int_{-\infty }^{\infty }G\left( x,y,t\right) \
\varphi \left( y\right) \ dy,\qquad \psi \left( x,0\right) =\varphi \left(
x\right)  \label{sum3}
\end{equation}%
for a suitable function $\varphi $ on $\boldsymbol{R};$ see Ref.~\cite%
{Suaz:Sus} for more details. Additional integrable cases are given with the
help of the gauge transformation.

Particular solutions of the corresponding nonlinear Schr\"{o}dinger
equations are obtained by the methods of Refs.~\cite{Cor-Sot:Lop:Sua:Sus}
and \cite{Cor-Sot:Sus}. A special case of the quantum parametric oscillator
with the Hamiltonian of the form (\ref{spc3}) is studied in detail. The
Green function is explicitly evaluated in terms of Airy functions by
equations (\ref{spc7})--(\ref{spc11}) and the corresponding transition
amplitudes are given in terms of a hypergeometric function by formula (\ref%
{spc20}). A discrete orthogonality relation for certain $_{2}F_{1}$
functions is derived from the fundamentals of quantum physics. It is
identified then as orthogonality property of special Meixner polynomials
with the help of a quadratic transformation. An extension to the general
case of parametric oscillator in quantum mechanics is also given. Relation
of the transition amplitudes with unitary irreducible representations of the
Lorentz group $SU\left( 1,1\right) $ is established. Further extension to
the quantum forced parametric oscillator is left to the reader.

We dedicate this paper to Professor Richard Askey on his 75th birthday for
his outstanding contributions to the area of classical analysis, special
functions and their numerous applications, and mathematical education.

\section{Appendix A: Solutions of Airy Equation}

Bessel functions are defined as%
\begin{equation}
J_{\nu }\left( z\right) =\left( \frac{z}{2}\right) ^{\nu }\sum_{k=0}^{\infty
}\frac{\left( -z^{2}/4\right) ^{k}}{k!\Gamma \left( \nu +k+1\right) }
\label{airy1}
\end{equation}%
and the modified Bessel functions are%
\begin{equation}
I_{\nu }\left( z\right) =\left( \frac{z}{2}\right) ^{\nu }\sum_{k=0}^{\infty
}\frac{\left( z^{2}/4\right) ^{k}}{k!\Gamma \left( \nu +k+1\right) }.
\label{airy2}
\end{equation}%
For an extensive theory of these functions, see Refs.~\cite{Ab:St}, \cite%
{An:As:Ro}, \cite{Ni:Uv}, \cite{Rain}, \cite{Sus:Trey}, \cite{Wa} and
references therein.

The Airy functions satisfy the second order differential equation%
\begin{equation}
u^{\prime \prime }-tu=0.  \label{airy3}
\end{equation}%
Their standard definitions are%
\begin{eqnarray}
Ai\left( t\right) &=&\frac{\sqrt{t}}{3}\left( I_{-1/3}\left( z\right)
-I_{1/3}\left( z\right) \right) ,  \label{airy4} \\
Bi\left( t\right) &=&\sqrt{\frac{t}{3}}\left( I_{-1/3}\left( z\right)
+I_{1/3}\left( z\right) \right)  \label{airy5}
\end{eqnarray}%
and%
\begin{eqnarray}
Ai\left( -t\right) &=&\frac{\sqrt{t}}{3}\left( J_{-1/3}\left( z\right)
+J_{1/3}\left( z\right) \right) ,  \label{airy6} \\
Bi\left( -t\right) &=&\sqrt{\frac{t}{3}}\left( J_{-1/3}\left( z\right)
-J_{1/3}\left( z\right) \right)  \label{airy7}
\end{eqnarray}%
with $z=\left( 2/3\right) t^{3/2}.$ The Wronskian is equal to%
\begin{equation}
W\left( Ai\left( t\right) ,Bi\left( t\right) \right) =\frac{1}{\pi }
\label{airy7a}
\end{equation}%
and the derivatives are given by%
\begin{eqnarray}
Ai^{\prime }\left( t\right) &=&\frac{\sqrt{t}}{3}\left( I_{2/3}\left(
z\right) -I_{-2/3}\left( z\right) \right) ,  \label{airy8} \\
Bi^{\prime }\left( t\right) &=&\sqrt{\frac{t}{3}}\left( I_{2/3}\left(
z\right) +I_{-2/3}\left( z\right) \right)  \label{airy9}
\end{eqnarray}%
and%
\begin{eqnarray}
Ai^{\prime }\left( -t\right) &=&\frac{t}{3}\left( J_{2/3}\left( z\right)
-J_{-2/3}\left( z\right) \right) ,  \label{airy10} \\
Bi^{\prime }\left( -t\right) &=&\frac{t}{\sqrt{3}}\left( J_{2/3}\left(
z\right) +J_{-2/3}\left( z\right) \right)  \label{airy11}
\end{eqnarray}%
with $z=\left( 2/3\right) t^{3/2}.$

In this paper we use the following pair of linearly independent solutions%
\begin{eqnarray}
a\left( t\right) &=&ai\left( t\right) =\frac{1}{3^{2/3}}\Gamma \left( \frac{1%
}{3}\right) t^{1/2}I_{1/3}\left( \frac{2}{3}t^{3/2}\right)  \label{airy12} \\
&=&t\sum_{k=0}^{\infty }\frac{\left( t^{3}/9\right) ^{k}}{k!\left(
4/3\right) _{k}}=\sum_{k=0}^{\infty }3^{k}\left( \frac{2}{3}\right) _{k}%
\frac{t^{3k+1}}{\left( 3k+1\right) !}  \notag \\
&=&t+\frac{t^{4}}{2^{2}3}+\frac{t^{7}}{2^{3}3^{2}7}+...  \notag
\end{eqnarray}%
and%
\begin{eqnarray}
b\left( t\right) &=&bi\left( t\right) =\frac{1}{3^{1/3}}\Gamma \left( \frac{2%
}{3}\right) t^{1/2}I_{-1/3}\left( \frac{2}{3}t^{3/2}\right)  \label{airy13}
\\
&=&\sum_{k=0}^{\infty }\frac{\left( t^{3}/9\right) ^{k}}{k!\left( 2/3\right)
_{k}}=\sum_{k=0}^{\infty }3^{k}\left( \frac{1}{3}\right) _{k}\frac{t^{3k}}{%
\left( 3k\right) !}  \notag \\
&=&1+\frac{t^{3}}{6}+\frac{t^{6}}{2^{2}3^{2}5}+...  \notag
\end{eqnarray}%
with $a\left( 0\right) =b^{\prime }\left( 0\right) =0,$ $a^{\prime }\left(
0\right) =b\left( 0\right) =1.$ Their relations with the standard Airy
functions $Ai\left( t\right) $ and $Bi\left( t\right) $ are%
\begin{equation}
\left( 
\begin{array}{c}
a\left( t\right) \\ 
b\left( t\right)%
\end{array}%
\right) =\frac{1}{2}\left( 
\begin{array}{cc}
-3^{1/3}\Gamma \left( 1/3\right) & 3^{-1/6}\Gamma \left( 1/3\right) \\ 
3^{2/3}\Gamma \left( 2/3\right) & 3^{1/6}\Gamma \left( 2/3\right)%
\end{array}%
\right) \left( 
\begin{array}{c}
Ai\left( t\right) \\ 
Bi\left( t\right)%
\end{array}%
\right)  \label{airy14aa}
\end{equation}%
with the inverse%
\begin{equation}
\left( 
\begin{array}{c}
Ai\left( t\right) \\ 
Bi\left( t\right)%
\end{array}%
\right) =\frac{1}{\pi }\left( 
\begin{array}{cc}
-3^{1/6}\Gamma \left( 2/3\right) & 3^{-1/6}\Gamma \left( 1/3\right) \\ 
3^{2/3}\Gamma \left( 2/3\right) & 3^{1/3}\Gamma \left( 1/3\right)%
\end{array}%
\right) \left( 
\begin{array}{c}
a\left( t\right) \\ 
b\left( t\right)%
\end{array}%
\right)  \label{airy14ab}
\end{equation}%
and the Wronskian is%
\begin{equation}
W\left( a\left( t\right) ,b\left( t\right) \right) =-1.  \label{airy14a}
\end{equation}%
The derivatives are given by%
\begin{eqnarray}
a^{\prime }\left( t\right) &=&ai^{\prime }\left( t\right) =\frac{1}{3^{2/3}}%
\Gamma \left( \frac{1}{3}\right) tI_{-2/3}\left( \frac{2}{3}t^{3/2}\right)
\label{airy14} \\
&=&\sum_{k=0}^{\infty }\frac{\left( t^{3}/9\right) ^{k}}{k!\left( 1/3\right)
_{k}}=\sum_{k=0}^{\infty }3^{k}\left( \frac{2}{3}\right) _{k}\frac{t^{3k}}{%
\left( 3k\right) !}  \notag \\
&=&1+\frac{t^{3}}{3}+\frac{t^{6}}{2^{3}3^{2}}+...  \notag
\end{eqnarray}%
and%
\begin{eqnarray}
b^{\prime }\left( t\right) &=&bi^{\prime }\left( t\right) =\frac{1}{3^{1/3}}%
\Gamma \left( \frac{2}{3}\right) tI_{2/3}\left( \frac{2}{3}t^{3/2}\right)
\label{airy15} \\
&=&\frac{t^{2}}{2}\sum_{k=0}^{\infty }\frac{\left( t^{3}/9\right) ^{k}}{%
k!\left( 5/3\right) _{k}}=\sum_{k=0}^{\infty }3^{k}\left( \frac{4}{3}\right)
_{k}\frac{t^{3k+2}}{\left( 3k+2\right) !}  \notag \\
&=&\frac{t^{2}}{2}+\frac{t^{5}}{2^{2}3\cdot 5}+...  \notag
\end{eqnarray}%
with the Wronskian%
\begin{equation}
W\left( a^{\prime }\left( t\right) ,b^{\prime }\left( t\right) \right) =t.
\label{airy16}
\end{equation}%
More facts about the Airy functions can be found in Refs.~\cite{Ab:St}, \cite%
{Ni:Uv}, and \cite{Olver}.

\section{Appendix B: Some Transformations of Hypergeometric Functions}

We derive the transformation formulas (\ref{spc21a}) as follows. In the even
case $k=2r$ and $n=2s,$ use the quadratic transformation \cite{An:As:Ro}, 
\cite{Rain}:%
\begin{equation}
_{2}F_{1}\left( 
\begin{array}{c}
a,\quad b \\ 
a+b+\dfrac{1}{2}%
\end{array}%
;\ 4z\left( 1-z\right) \right) =\ _{2}F_{1}\left( 
\begin{array}{c}
2a,\quad 2b \\ 
a+b+\dfrac{1}{2}%
\end{array}%
;\ z\right)  \label{quad}
\end{equation}%
followed by a transformation:%
\begin{equation}
\ _{2}F_{1}\left( 
\begin{array}{c}
a,\ -r \\ 
c%
\end{array}%
;\ z\right) =\frac{\left( c-a\right) _{r}}{\left( c\right) _{r}}\ \
_{2}F_{1}\left( 
\begin{array}{c}
a,\quad -r \\ 
1+a-c-r%
\end{array}%
;\ 1-z\right) ,  \label{T2107}
\end{equation}%
where $\left( a\right) _{r}=a\left( a+1\right) ...\left( a+r-1\right)
=\Gamma \left( a+r\right) /\Gamma \left( a\right) ,$ for the terminating
hypergeometric function. The reflection formula for gamma function:%
\begin{equation}
\Gamma \left( z\right) \Gamma \left( 1-z\right) =\frac{\pi }{\sin \pi z}
\label{gammaref}
\end{equation}%
allows to complete the proof.

In the odd case $k=2r+1$ and $n=2s+1,$ one can use the quadratic
transformation (\ref{quad}) for pure imaginary values of $\zeta $ with $%
0<\left( \func{Im}\zeta \right) ^{2}<2,$ when the series converges. Apply
the familiar transformation \cite{An:As:Ro}, \cite{Ni:Uv}, \cite{Rain}:%
\begin{equation}
\ _{2}F_{1}\left( 
\begin{array}{c}
a,\ b \\ 
c%
\end{array}%
;\ z\right) =\left( 1-z\right) ^{c-a-b}\ _{2}F_{1}\left( 
\begin{array}{c}
c-a,\quad c-b \\ 
c%
\end{array}%
;\ z\right)  \label{T2103}
\end{equation}%
to get back to a terminating hypergeometric function. The end result, namely,%
\begin{eqnarray}
&&_{2}F_{1}\left( 
\begin{array}{c}
-2r-1,\ -2s-1\medskip \\ 
-r-s-1/2%
\end{array}%
;\ \dfrac{1}{2}\left( 1+i\zeta \right) \right)  \label{appb} \\
&&\quad =\ _{2}F_{1}\left( 
\begin{array}{c}
-r-1/2,\ -s-1/2\medskip \\ 
-r-s-1/2%
\end{array}%
;\ 1+\zeta ^{2}\right)  \notag \\
&&\quad \quad =-i\zeta \ _{2}F_{1}\left( 
\begin{array}{c}
-r,\ -s\medskip \\ 
-r-s-1/2%
\end{array}%
;\ 1+\zeta ^{2}\right) ,  \notag
\end{eqnarray}%
is valid by analytic continuation in the entire complex plane. Our choice of
the branch of the radical corresponds to the correct special value at $%
r=s=0. $ Use the transformation (\ref{T2107}) and reflection formula (\ref%
{gammaref}) in order to complete the proof.

The Clausen formula \cite{An:As:Ro}:%
\begin{equation}
\left[ _{2}F_{1}\left( 
\begin{array}{c}
a,\quad b \\ 
a+b+\dfrac{1}{2}%
\end{array}%
;\ z\right) \right] ^{2}=\ _{3}F_{2}\left( 
\begin{array}{c}
2a,\quad 2b,\quad a+b \\ 
a+b+\dfrac{1}{2},\quad 2a+2b%
\end{array}%
;\ z\right)  \label{claus}
\end{equation}%
and the duplication formula for gamma function:%
\begin{equation}
\Gamma \left( 2z\right) =\frac{2^{2z-1}}{\sqrt{\pi }}\Gamma \left( z\right)
\Gamma \left( z+\frac{1}{2}\right)  \label{gammadouble}
\end{equation}%
have been used in sections~8 and 9.

\noindent \textbf{Acknowledgment.\/} We thank Professor Richard Askey for
motivation, valuable discussions and encouragement.

\end{document}